\expandafter\edef\csname hypers@fe\endcsname{\catcode
                                             `\noexpand @=\the\catcode`\@}%
\catcode`\@=11
%
% Check if the file is already included
%
\ifx\hyperd@ne\hyper@ndefined
 \global\let\hyperd@ne=\relax
\else
 \errhelp{hyperbasics.tex needs to be included only once outside
          of any {...} or \begingroup...\endgroup. You have tried to
          include it more than once. If the previous include was indeed
          outside any groupings, continue and all will be well.}%
 \errmessage{Input this file only once!}%
  
\fi
%
% Version number
%
\def\hyperv@rsion{8}%
%
% Check and input a previous .hrf file if it exists
%
\newread\hyperf@le
\def\hyperf@lename{\jobname.hrf}%
\immediate\openin\hyperf@le\hyperf@lename\relax
\ifeof\hyperf@le\relax
 \immediate\closein\hyperf@le\relax
\else
 \immediate\closein\hyperf@le\relax
 \input \hyperf@lename
\fi
%
% Open a new .hrf file
%
\newwrite\hyperf@le
\immediate\openout\hyperf@le\hyperf@lename
%%%%
% MAIN SECTION
%%%%
%
% define a token register
%
\newtoks\hypert@ks
%
% Define a convenient macro to hold the character #
%
\edef\hypert@mp{\catcode`\noexpand\#=\the\catcode`\#}%
\catcode`\#=12
\def\hyperh@sh{#}%
\hypert@mp
\let\hypert@mp=\relax
\let\hyper@nd=\relax
\def\hyperstr@pquote"#1"#2\hyper@nd{\ifx\hyper@ndefined#2\hyper@ndefined#1\else
                                    \ifx\hyper@ndefined#1\hyper@ndefined
                                    \hyperstr@pquote#2"\hyper@nd\else
                                    #1\hyperstr@pquote"#2"\hyper@nd\fi\fi}%
\def\hyperstr@pblank" #1 #2\hyper@nd"{\ifx\hyper@ndefined#2\hyper@ndefined#1\else
                                    \ifx\hyper@ndefined#1\hyper@ndefined
                                    \hyperstr@pblank"#2 \hyper@nd"\else
                                    #1\hyperstr@pblank" #2 \hyper@nd"\fi\fi}
\long\def\hyper@nchor#1#2{\edef\hyperm@cro{html:<A #1>}%
                          \special\expandafter{\hyperm@cro}%
                          {#2}}%
\def\hyper@atm@ning#1->#2\hyper@nd{#2}
\def\hyperlink#1{\edef\hypert@mp{#1}%
               \edef\hypert@mp{\expandafter\hyper@atm@ning\meaning\hypert@mp
                               \hyper@nd}%
               \edef\hypert@mp"{ \expandafter\hyperstr@pquote\expandafter"%
                               \hypert@mp"\hyper@nd}%
               \edef\hypert@mp{\expandafter\hyperstr@pblank\expandafter%
                               "\hypert@mp" \hyper@nd"}%
               \hyper@nchor{href=\expandafter"\hypert@mp"}}%
\def\hypertarget#1{\edef\hypert@mp{#1}%
               \edef\hypert@mp{\expandafter\hyper@atm@ning\meaning\hypert@mp
                               \hyper@nd}%
               \edef\hypert@mp"{ \expandafter\hyperstr@pquote\expandafter"%
                               \hypert@mp"\hyper@nd}%
               \edef\hypert@mp{\expandafter\hyperstr@pblank\expandafter%
                               "\hypert@mp" \hyper@nd"}%
               \hyper@nchor{name=\expandafter"\hypert@mp"}}%
\def\hyperref{\afterassignment\hyperr@f\let\hyperp@ram}
\def\hyperr@f{\ifx\hyperp@ram{\iffalse}\fi
               \expandafter\expandafter\expandafter\hyperr@@
               \expandafter{%
              \else
               \iffalse}\fi
               \ifx\hyperp@ram\hyper@ndefined
                 \message{Undefined reference}%
                 \def\hyperp@r@m{{}{undefined}{}}%
               \else
                 \edef\hyperp@r@m{\hyperp@ram}%
               \fi
               \expandafter\expandafter\expandafter\hyperr@@
               \expandafter\hyperp@r@m
              \fi}%
% refer to #1, \hyperh@sh#2.#3 or #1\hyperh@sh#2.#3
% depending on what is blank/nonblank
\def\hyperr@@#1#2#3{\ifx\hyper@ndefined#1\hyper@ndefined
                    \hypert@ks\expandafter{\hyperh@sh#2.#3}%
                    \else
                     \ifx\hyper@ndefined#2#3\hyper@ndefined
                      \hypert@ks{#1}%
                     \else
                      \def\hypert@mp{#1}%
                      \hypert@ks\expandafter\expandafter\expandafter
                      {\expandafter\hypert@mp\hyperh@sh#2.#3}%
                     \fi
                    \fi
                    \expandafter\hyperlink\expandafter{\the\hypert@ks}}%
\def\hyperdef#1#2#3{{\global\escapechar=`\\\relax
                     \edef\hypert@mp{\hyperstr@pquote"#2.#3"\hyper@nd}%
                     \expandafter\ifx\csname hyperd@\meaning\hypert@mp
                     \endcsname
                     \relax
                     \expandafter\gdef\csname hyperd@\meaning\hypert@mp
                     \endcsname{}%
                     \gdef#1{{}{\hyperstr@pquote"#2"\hyper@nd}%
                               {\hyperstr@pquote"#3"\hyper@nd}}%
                     \immediate\write\hyperf@le{\def\noexpand#1{#1}}%
                     \xdef\hypert@mp{\global\let\noexpand\hypert@mp=\relax
                                     \noexpand\hypertarget{\hypert@mp}}%
                     \global\hypert@ks={\hypert@mp}%
                     \else
                     \message\expandafter{'\hypert@mp' duplicate}%
                     \global\let\hypert@mp=\relax
                     \global\hypert@ks={\hyperdef{#1}{#2}{#3@}}%
                     \fi}\the\hypert@ks}%

\def\hyper@nique#1#2#3#4{\global\escapechar=`\\\relax
                     \edef\hypert@mp{\hyperstr@pquote"#2.#3"\hyper@nd}%
                     \expandafter\ifx\csname hyperd@\meaning\hypert@mp
                     \endcsname
                     \relax
                     \gdef#1{{}{\hyperstr@pquote"#2"\hyper@nd}%
                               {\hyperstr@pquote"#3"\hyper@nd}}%
                     \global\let\hypert@mp=\relax
                     #4%
                     \else
                     \global\let\hypert@mp=\relax
                     \hyper@nique{#1}{#2}{#3@}{#4}%
                     \fi
                     }%

%%% 
% protection macros
%%%
\let\hyper@@@@=\relax
\def\hyper@@{\let\hyper@@@=\relax}%
\hyper@@
\def\hyper@{\relax\let\hyper@@@\noexpand\hyper@\noexpand}%
\def\hyperpr@ref{\hyper@@\hyperref}
\def\hyperpr@def{\hyper@@\hyperdef}

% As per pg's suggestion
\let\href\hyperlink

%
% Restore the catcode of @
%
\hypers@fe

\catcode`\@=11
%%% saclay A4 paper:
\def\unredoffs{\voffset=13mm \hoffset=6.5truemm} 
\def\redoffs{\voffset=-12.truemm\hoffset=-3truemm} 

\newif\ifbookmode
\bookmodefalse
%
% ****************************** INDEX
\newwrite\inx
\def\book{\bookmodetrue\immediate\openout\inx=\jobname.inx%
 \hsize=120mm\vsize=195mm} 
\def\@#1{\noindent\ifbookmode\write\inx{#1,\space
\number\pageno.\par}\fi}

%---------------------------------------------------------------------%
\newbox\leftpage \newdimen\fullhsize \newdimen\hstitle \newdimen\hsbody
\newdimen\hdim
%\tolerance=1000
\hfuzz=1pt
\ifx\hyperdef\UNd@FiNeD\def\hyperdef#1#2#3#4{#4}\def\hyperref#1#2#3#4{#4}\fi
%%%%%%%%%%%%%%%%%%%%%%%%%%%%%%%%%%%%%%%%%%%%%%%%%
\def\bigans{b }
%\message{ big or little (b/l)? }\read-1 to\answ
\def\answ{b }
\ifx\answ\bigans\message{(Format simple colonne 12pts.}
\magnification=1200 \unredoffs%\ifbookmode \hsize=144truemm\vsize=234truemm%
%\else
\hsize=122.5mm\vsize=182.5mm
%\fi   
\hsbody=\hsize \hstitle=\hsize %take default values for unreduced format
\else\message{(Format simple colonne, 10pts.} \let\l@r=L
\magnification=1000 
\redoffs%
\hsize=122.5mm\vsize=182.5mm
\hsbody=\hsize \hstitle=\hsize %take default values for unreduced format
\fi
% 
%%%%%%%%%%%%%%%%%%%%%% fonts, Dirac slash %%%%%%%%%%%%%%%
% 5/12/02  twbfx added,
% 5/12/02   \chapfnt=cmbx10 scaled 1440 \font\headbf=cmbx9
% 30/04/07 adddition of chapfonts
% 25/02/09 MACRO \Bfg for bold lowercase greek letters 
% 22/04/09 \skewchar for \Bfg

\def\sla#1{\mkern-1.5mu\raise0.4pt\hbox{$\not$}\mkern1.2mu #1\mkern 0.7mu}
\def\Dbar{\mkern-1.5mu\raise0.4pt\hbox{$\not$}\mkern-.1mu {\rm D}\mkern.1mu}
\def\Abar{\mkern1.mu\raise0.4pt\hbox{$\not$}\mkern-1.3mu A\mkern.1mu}
\def\Bbar{\mkern-0.mu\raise0.4pt\hbox{$\not$}\mkern-.3mu B\mkern 0.6mu}
\newskip\tableskipamount \tableskipamount=8pt plus 3pt minus 3pt

%****************************
%%%% chapters
\newdimen\chapskip
\chapskip=17.5mm
\font\twbfx=cmbx10 scaled 1200
\font\chapfnt=cmbx10 scaled 1440
\font\chapbxten=cmbx10
\font\chapbxseven=cmbx7

\font\ssbx=cmssbx10  

\font\chaprm=cmr10 scaled 1440
\font\chaprmscript=cmr10
\font\chaprmseven=cmr7
\font\chapssfnt=cmssbx10 scaled 1440

\font\chapibfnt=cmmib10 scaled 1440
\font\chapmifnt=cmmi10 scaled 1440
\font\chapsyfnt=cmsy10 scaled 1440
\font\chapexfnt=cmex10 scaled 1440
\font\chapibten=cmmib10
\font\chapibseven=cmmib7
\font\chapmiscript=cmmi10  
\font\chapsyscript=cmsy10  
\font\chapexscript=cmex10
\font\chapmiseven=cmmi7
\font\chapsyseven=cmsy7
\font\chapexseven=cmex7  
%%%%%%%%%%%%%%%%%%%%%%%%%%
\def\chapfont{
%\textfont0=\chapssfnt \scriptfont0=\chapssten \scriptscriptfont0=\chapssseven
%\def\rm{\fam0\chapssfnt}
\textfont0=\chaprm\scriptfont0=\chaprmscript\scriptscriptfont0=\chaprmseven
\textfont1=\chapmifnt \scriptfont1=\chapmiscript  \scriptscriptfont1=\chapmiseven
\textfont2=\chapsyfnt \scriptfont2=\chapsyscript\scriptscriptfont2=\chapsyseven
\textfont3=\chapexfnt \scriptfont3=\chapexscript \scriptscriptfont3=\chapexseven
%\textfont\itfam=\capit \def\it{\fam\itfam\capit} % \it is family 4
%\textfont\slfam=\capsl  \def\sl{\fam\slfam\capsl} % \sl is family 5
\textfont\bffam=\chapfnt \scriptfont\bffam=\chapbxten
\scriptscriptfont\bffam=\chapbxseven
\def\bf{\fam\bffam\chapfnt} % \bf is family 6
\textfont4=\chapibfnt \scriptfont4=\chapibten \scriptscriptfont4=\chapibseven
\abovedisplayskip=17pt plus 5pt minus 13pt
\belowdisplayskip=\abovedisplayskip
%\smallskipamount=2.7pt plus 1pt minus 1pt
%\medskipamount=5.4pt plus 2pt minus 2pt
%\bigskipamount=10.8pt plus 3.6pt minus 3.6pt
\normalbaselineskip=17pt
\setbox\strutbox=\hbox{\vrule height12.2pt depth5.0pt width0pt}
\normalbaselines \chapssfnt}

\font\caprm=cmr9
\font\capit=cmti9
\font\capbf=cmbx9
\font\capsl=cmsl9
\font\capmi=cmmi9
\font\capex=cmex9
\font\capsy=cmsy9

\def\makeheadline{\vbox to 0pt{\vskip-22.5pt
\line{\vbox to8.5pt{}\the\headline}\vss}\nointerlineskip}
%%%%%%%%%%%%%%%%%%%%%%%%%%%%%%%%%%%%%%%%%%%%%%%%%%%%%%%%%%%%%%%%%%%%%%%%
\font\headrm=cmr10

%****************************
\font\sixrm=cmr6
\font\fiverm=cmr5
\font\sixmi=cmmi6
\font\fivemi=cmmi5
\font\sixsy=cmsy6
\font\fivesy=cmsy5
\font\sixbf=cmbx6
\font\fivebf=cmbx5
\skewchar\capmi='177 \skewchar\sixmi='177 \skewchar\fivemi='177
\skewchar\capsy='60 \skewchar\sixsy='60 \skewchar\fivesy='60

% *************************************************************************

% ****************************************************************************
%		*****	  MSSYMB.TeX	*****		       20 Sept 91
%
%	This file contains the definitions for the symbols in the two
%	"extra symbols" fonts created at the American Math. Society.
%
%       The old fonts msxm et msym have been replaced by msam et msbm. 

\catcode`\@=11
%%%%%%%%%%%%%%%%%%%%%%%%%%%%%%%%%%%%%%%
%***************************************************
  %obsolete??
\font\tenbi=cmmib10 
\font\ninebi=cmmib9
\font\sevenbi=cmmib7 
\font\fivebi=cmmib5
\textfont4=\tenbi \scriptfont4=\sevenbi \scriptscriptfont4=\fivebi
\newfam\mibfam
\textfont\mibfam=\tenbi \scriptfont\mibfam=\sevenbi \scriptscriptfont\mibfam=\fivebi
\def\Bfg{\ifmmode\let\next\Bfg@\else
 \def\next{\errmessage{Use \string\Bfg\space only in math mode}}\fi\next}
\def\Bfg@#1{{\Bfg@@{#1}}}
\def\Bfg@@#1{\fam\mibfam#1}
\skewchar\tenbi='177\skewchar\sevenbi='177
%%%%%%%%%%%%%%%%%%%%%%%%%%%%%%%%%%%%%%%%%%%%%%%%%%%%%%%%%%%%%%
\mathchardef\alpha="710B
\mathchardef\beta="710C 
\mathchardef\gamma="710D
\mathchardef\delta="710E
\mathchardef\epsilon="710F  
\mathchardef\zeta="7110  
\mathchardef\eta="7111 
\mathchardef\theta="7112 
\mathchardef\iota="7113 
\mathchardef\kappa="7114 
\mathchardef\lambda="7115 
\mathchardef\mu="7116 
\mathchardef\nu="7117 
\mathchardef\xi="7118 
\mathchardef\pi="7119 
\mathchardef\rho="711A  
\mathchardef\sigma="711B 
\mathchardef\tau="711C 
\mathchardef\upsilon="711D 
\mathchardef\phi="711E 
\mathchardef\chi="711F 
\mathchardef\psi="7120 
\mathchardef\omega="7121 
\mathchardef\varepsilon="7122 
\mathchardef\vartheta="7123
\mathchardef\varpi="7124 
\mathchardef\varrho="7125 
\mathchardef\varsigma="7126 
\mathchardef\varphi="7127
%%%%%%%%%%%%%%%%%%%%%%%% OBSOLETE except for titles %%%%%%%%%%%%%%%%%%%%% %%%%%%%%%%%%%%%%%%%%%%%%%%%%%%%%%%%%%%%%%%%%%%%%%%%%%%%%%%%%%%%%%%%%%%%%%
\mathchardef\alphab="040B
\mathchardef\betab="040C 
\mathchardef\gammab="040D
\mathchardef\deltab="040E
\mathchardef\epsilonb="040F  
\mathchardef\zetab="0410  
\mathchardef\etab="0411 
\mathchardef\thetab="0412 
\mathchardef\iotab="0413 
\mathchardef\kappab="0414 
\mathchardef\lambdab="0415 
\mathchardef\mub="0416 
\mathchardef\nub="0417 
\mathchardef\xib="0418 
\mathchardef\pib="0419 
\mathchardef\rhob="041A  
\mathchardef\sigmab="041B 
\mathchardef\taub="041C 
\mathchardef\upsilonb="041D 
\mathchardef\phib="041E 
\mathchardef\chib="041F 
\mathchardef\psib="0420 
\mathchardef\omegab="0421 
\mathchardef\varepsilonb="0422 
\mathchardef\varthetab="0423
\mathchardef\varpib="0424 
\mathchardef\varrhob="0425 
\mathchardef\varsigmab="0426 
\mathchardef\varphib="0427 
 %%%%%%%%%%%%%%%%%%%%%%%%%%%%%%%%%%%%%%%%%%%%%%%%%%%%%%%%%%%%
\font\tenmsa=msam10
\font\sevenmsa=msam7
\font\fivemsa=msam5
\font\tenmsb=msbm10
\font\sevenmsb=msbm7
\font\fivemsb=msbm5
\newfam\msafam
\newfam\msbfam
\textfont\msafam=\tenmsa  \scriptfont\msafam=\sevenmsa
  \scriptscriptfont\msafam=\fivemsa
\textfont\msbfam=\tenmsb  \scriptfont\msbfam=\sevenmsb
  \scriptscriptfont\msbfam=\fivemsb

\def\hexnumber@#1{\ifcase#1 0\or1\or2\or3\or4\or5\or6\or7\or8\or9\or
	A\or B\or C\or D\or E\or F\fi }
%  The following 13 lines establish the use of the Euler Fraktur font.
%  To use this font, remove % from beginning of these lines.
\font\teneuf=eufm10
\font\seveneuf=eufm7
\font\fiveeuf=eufm5
\newfam\euffam
\textfont\euffam=\teneuf
\scriptfont\euffam=\seveneuf
\scriptscriptfont\euffam=\fiveeuf
\def\frak{\ifmmode\let\next\frak@\else
 \def\next{\Err@{Use \string\frak\space only in math mode}}\fi\next}
\def\goth{\ifmmode\let\next\frak@\else
 \def\next{\Err@{Use \string\goth\space only in math mode}}\fi\next}
\def\frak@#1{{\frak@@{#1}}}
\def\frak@@#1{\fam\euffam#1}
%  End definition of Euler Fraktur font.

\edef\msa@{\hexnumber@\msafam}
\edef\msb@{\hexnumber@\msbfam}

\mathchardef\boxdot="2\msa@00
\mathchardef\boxplus="2\msa@01
\mathchardef\boxtimes="2\msa@02
\mathchardef\square="0\msa@03
\mathchardef\blacksquare="0\msa@04
\mathchardef\centerdot="2\msa@05
\mathchardef\lozenge="0\msa@06
\mathchardef\blacklozenge="0\msa@07
\mathchardef\circlearrowright="3\msa@08
\mathchardef\circlearrowleft="3\msa@09
\mathchardef\rightleftharpoons="3\msa@0A
\mathchardef\leftrightharpoons="3\msa@0B
\mathchardef\boxminus="2\msa@0C
\mathchardef\Vdash="3\msa@0D
\mathchardef\Vvdash="3\msa@0E
\mathchardef\vDash="3\msa@0F
\mathchardef\twoheadrightarrow="3\msa@10
\mathchardef\twoheadleftarrow="3\msa@11
\mathchardef\leftleftarrows="3\msa@12
\mathchardef\rightrightarrows="3\msa@13
\mathchardef\upuparrows="3\msa@14
\mathchardef\downdownarrows="3\msa@15
\mathchardef\upharpoonright="3\msa@16

\mathchardef\downharpoonright="3\msa@17
\mathchardef\upharpoonleft="3\msa@18
\mathchardef\downharpoonleft="3\msa@19
\mathchardef\rightarrowtail="3\msa@1A
\mathchardef\leftarrowtail="3\msa@1B
\mathchardef\leftrightarrows="3\msa@1C
\mathchardef\rightleftarrows="3\msa@1D
\mathchardef\Lsh="3\msa@1E
\mathchardef\Rsh="3\msa@1F
\mathchardef\rightsquigarrow="3\msa@20
\mathchardef\leftrightsquigarrow="3\msa@21
\mathchardef\looparrowleft="3\msa@22
\mathchardef\looparrowright="3\msa@23
\mathchardef\circeq="3\msa@24
\mathchardef\succsim="3\msa@25
\mathchardef\gtrsim="3\msa@26
\mathchardef\gtrapprox="3\msa@27
\mathchardef\multimap="3\msa@28
\mathchardef\therefore="3\msa@29
\mathchardef\because="3\msa@2A
\mathchardef\doteqdot="3\msa@2B

\mathchardef\triangleq="3\msa@2C
\mathchardef\precsim="3\msa@2D
\mathchardef\lesssim="3\msa@2E
\mathchardef\lessapprox="3\msa@2F
\mathchardef\eqslantless="3\msa@30
\mathchardef\eqslantgtr="3\msa@31
\mathchardef\curlyeqprec="3\msa@32
\mathchardef\curlyeqsucc="3\msa@33
\mathchardef\preccurlyeq="3\msa@34
\mathchardef\leqq="3\msa@35
\mathchardef\leqslant="3\msa@36
\mathchardef\lessgtr="3\msa@37
\mathchardef\backprime="0\msa@38
\mathchardef\risingdotseq="3\msa@3A
\mathchardef\fallingdotseq="3\msa@3B
\mathchardef\succcurlyeq="3\msa@3C
\mathchardef\geqq="3\msa@3D
\mathchardef\geqslant="3\msa@3E
\mathchardef\gtrless="3\msa@3F
\mathchardef\sqsubset="3\msa@40
\mathchardef\sqsupset="3\msa@41
\mathchardef\vartriangleright="3\msa@42
\mathchardef\vartriangleleft="3\msa@43
\mathchardef\trianglerighteq="3\msa@44
\mathchardef\trianglelefteq="3\msa@45
\mathchardef\bigstar="0\msa@46
\mathchardef\between="3\msa@47
\mathchardef\blacktriangledown="0\msa@48
\mathchardef\blacktriangleright="3\msa@49
\mathchardef\blacktriangleleft="3\msa@4A
\mathchardef\vartriangle="0\msa@4D
\mathchardef\blacktriangle="0\msa@4E
\mathchardef\triangledown="0\msa@4F
\mathchardef\eqcirc="3\msa@50
\mathchardef\lesseqgtr="3\msa@51
\mathchardef\gtreqless="3\msa@52
\mathchardef\lesseqqgtr="3\msa@53
\mathchardef\gtreqqless="3\msa@54
\mathchardef\Rrightarrow="3\msa@56
\mathchardef\Lleftarrow="3\msa@57
\mathchardef\veebar="2\msa@59
\mathchardef\barwedge="2\msa@5A
\mathchardef\doublebarwedge="2\msa@5B
\mathchardef\angle="0\msa@5C
\mathchardef\measuredangle="0\msa@5D
\mathchardef\sphericalangle="0\msa@5E
\mathchardef\varpropto="3\msa@5F
\mathchardef\smallsmile="3\msa@60
\mathchardef\smallfrown="3\msa@61
\mathchardef\Subset="3\msa@62
\mathchardef\Supset="3\msa@63
\mathchardef\Cup="2\msa@64

\mathchardef\Cap="2\msa@65

\mathchardef\curlywedge="2\msa@66
\mathchardef\curlyvee="2\msa@67
\mathchardef\leftthreetimes="2\msa@68
\mathchardef\rightthreetimes="2\msa@69
\mathchardef\subseteqq="3\msa@6A
\mathchardef\supseteqq="3\msa@6B
\mathchardef\bumpeq="3\msa@6C
\mathchardef\Bumpeq="3\msa@6D
\mathchardef\lll="3\msa@6E

\mathchardef\ggg="3\msa@6F

\mathchardef\circledS="0\msa@73
\mathchardef\pitchfork="3\msa@74
\mathchardef\dotplus="2\msa@75
\mathchardef\backsim="3\msa@76
\mathchardef\backsimeq="3\msa@77
\mathchardef\complement="0\msa@7B
\mathchardef\intercal="2\msa@7C
\mathchardef\circledcirc="2\msa@7D
\mathchardef\circledast="2\msa@7E
\mathchardef\circleddash="2\msa@7F
\def\ulcorner{\delimiter"4\msa@70\msa@70 }
\def\urcorner{\delimiter"5\msa@71\msa@71 }
\def\llcorner{\delimiter"4\msa@78\msa@78 }
\def\lrcorner{\delimiter"5\msa@79\msa@79 }
\def\yen{\mathhexbox\msa@55 }
\def\checkmark{\mathhexbox\msa@58 }
\def\circledR{\mathhexbox\msa@72 }
\def\maltese{\mathhexbox\msa@7A }
\mathchardef\lvertneqq="3\msb@00
\mathchardef\gvertneqq="3\msb@01
\mathchardef\nleq="3\msb@02
\mathchardef\ngeq="3\msb@03
\mathchardef\nless="3\msb@04
\mathchardef\ngtr="3\msb@05
\mathchardef\nprec="3\msb@06
\mathchardef\nsucc="3\msb@07
\mathchardef\lneqq="3\msb@08
\mathchardef\gneqq="3\msb@09
\mathchardef\nleqslant="3\msb@0A
\mathchardef\ngeqslant="3\msb@0B
\mathchardef\lneq="3\msb@0C
\mathchardef\gneq="3\msb@0D
\mathchardef\npreceq="3\msb@0E
\mathchardef\nsucceq="3\msb@0F
\mathchardef\precnsim="3\msb@10
\mathchardef\succnsim="3\msb@11
\mathchardef\lnsim="3\msb@12
\mathchardef\gnsim="3\msb@13
\mathchardef\nleqq="3\msb@14
\mathchardef\ngeqq="3\msb@15
\mathchardef\precneqq="3\msb@16
\mathchardef\succneqq="3\msb@17
\mathchardef\precnapprox="3\msb@18
\mathchardef\succnapprox="3\msb@19
\mathchardef\lnapprox="3\msb@1A
\mathchardef\gnapprox="3\msb@1B
\mathchardef\nsim="3\msb@1C
%\mathchardef\napprox="3\msb@1D
\mathchardef\ncong="3\msb@1D

\mathchardef\varsubsetneq="3\msb@20
\mathchardef\varsupsetneq="3\msb@21
\mathchardef\nsubseteqq="3\msb@22
\mathchardef\nsupseteqq="3\msb@23
\mathchardef\subsetneqq="3\msb@24
\mathchardef\supsetneqq="3\msb@25
\mathchardef\varsubsetneqq="3\msb@26
\mathchardef\varsupsetneqq="3\msb@27
\mathchardef\subsetneq="3\msb@28
\mathchardef\supsetneq="3\msb@29
\mathchardef\nsubseteq="3\msb@2A
\mathchardef\nsupseteq="3\msb@2B
\mathchardef\nparallel="3\msb@2C
\mathchardef\nmid="3\msb@2D
\mathchardef\nshortmid="3\msb@2E
\mathchardef\nshortparallel="3\msb@2F
\mathchardef\nvdash="3\msb@30
\mathchardef\nVdash="3\msb@31
\mathchardef\nvDash="3\msb@32
\mathchardef\nVDash="3\msb@33
\mathchardef\ntrianglerighteq="3\msb@34
\mathchardef\ntrianglelefteq="3\msb@35
\mathchardef\ntriangleleft="3\msb@36
\mathchardef\ntriangleright="3\msb@37
\mathchardef\nleftarrow="3\msb@38
\mathchardef\nrightarrow="3\msb@39
\mathchardef\nLeftarrow="3\msb@3A
\mathchardef\nRightarrow="3\msb@3B
\mathchardef\nLeftrightarrow="3\msb@3C
\mathchardef\nleftrightarrow="3\msb@3D
\mathchardef\divideontimes="2\msb@3E
\mathchardef\varnothing="0\msb@3F
\mathchardef\nexists="0\msb@40
\mathchardef\mho="0\msb@66
\mathchardef\eth="0\msb@67
\mathchardef\eqsim="3\msb@68
\mathchardef\beth="0\msb@69
\mathchardef\gimel="0\msb@6A
\mathchardef\daleth="0\msb@6B
\mathchardef\lessdot="3\msb@6C
\mathchardef\gtrdot="3\msb@6D
\mathchardef\ltimes="2\msb@6E
\mathchardef\rtimes="2\msb@6F
\mathchardef\shortmid="3\msb@70
\mathchardef\shortparallel="3\msb@71
\mathchardef\smallsetminus="2\msb@72
\mathchardef\thicksim="3\msb@73
\mathchardef\thickapprox="3\msb@74
\mathchardef\approxeq="3\msb@75
\mathchardef\succapprox="3\msb@76
\mathchardef\precapprox="3\msb@77
\mathchardef\curvearrowleft="3\msb@78
\mathchardef\curvearrowright="3\msb@79
\mathchardef\digamma="0\msb@7A
\mathchardef\varkappa="0\msb@7B
\mathchardef\hslash="0\msb@7D
\mathchardef\hbar="0\msb@7E
\mathchardef\backepsilon="3\msb@7F
\def\Bbb{\ifmmode\let\next\Bbb@\else
 \def\next{\errmessage{Use \string\Bbb\space only in math mode}}\fi\next}
\def\Bbb@#1{{\Bbb@@{#1}}}
\def\Bbb@@#1{\fam\msbfam#1}
%%%%%%%%%%%%%%%%%%%%%%%%%%%%%%%%%%%%%%%%%%%%%%%%%%%%%%%%%%%%%

\def\elevenpoint{
\textfont0=\caprm \scriptfont0=\sixrm \scriptscriptfont0=\fiverm
\def\rm{\fam0\caprm}
\textfont1=\capmi \scriptfont1=\sixmi \scriptscriptfont1=\fivemi
\textfont2=\capsy \scriptfont2=\sixsy \scriptscriptfont2=\fivesy
\textfont3=\capex \scriptfont3=\capex \scriptscriptfont3=\capex
\textfont\itfam=\capit \def\it{\fam\itfam\capit} % \it is family 4
\textfont\slfam=\capsl  \def\sl{\fam\slfam\capsl} % \sl is family 5
\textfont\bffam=\capbf \scriptfont\bffam=\sixbf
\scriptscriptfont\bffam=\fivebf
\def\bf{\fam\bffam\capbf} % \bf is family 6
\textfont4=\ninebi \scriptfont4=\sevenbi \scriptscriptfont4=\fivebi
\abovedisplayskip=11pt plus 3pt minus 8pt
\belowdisplayskip=\abovedisplayskip
\smallskipamount=2.7pt plus 1pt minus 1pt
\medskipamount=5.4pt plus 2pt minus 2pt
\bigskipamount=10.8pt plus 3.6pt minus 3.6pt
\normalbaselineskip=11pt
\setbox\strutbox=\hbox{\vrule height7.8pt depth3.2pt width0pt}
\normalbaselines \rm}
\catcode`\@=12
\def\sla#1{\mkern-1.5mu\raise0.4pt\hbox{$\not$}\mkern1.2mu #1\mkern 0.7mu}
\def\Dbar{\mkern-1.5mu\raise0.4pt\hbox{$\not$}\mkern-.1mu {\rm D}\mkern.1mu}
\def\Abar{\mkern1.mu\raise0.4pt\hbox{$\not$}\mkern-1.3mu A\mkern.1mu}
\nopagenumbers
%%%%%%%%%%%%%%%%%% headline %%%%%%%%%%%%%%%%%%%%%%%% 
\def\makeheadline{\vbox to 0pt{\vskip-27pt
\line{\vbox to8.5pt{}\the\headline}\vss}\nointerlineskip}
\def\subsectionname{}
\def\sectionname{}
\headline={\ifnum\pageno=1\hfill\else\draftdate\hfil{\headrm\folio}%
\hfil\hphantom{\draftdate}\fi}	 

%%%%%%%%%%%%%%%%%%%%%%%%%%%%%%%%%%%%%%%%%%%%%%%% %********* end ouput macros
%%%%%%%%%%%%%%%%%%%%%%%%%%%%%%%%%%%%%%%%%%%%%%%% 
% ****** extrait de definit.tex (obsolete ?)

% **************************************************************
\newcount\yearltd\yearltd=\year\advance\yearltd by -2000
\newif\ifdraftmode
\draftmodefalse
\def\draft{\draftmodetrue{\count255=\time\divide\count255 by 60
\xdef\hourmin{\number\count255} 
  \multiply\count255 by-60\advance\count255 by\time
  \xdef\hourmin{\hourmin:\ifnum\count255<10 0\fi\the\count255}}}
\def\draftdate{\ifdraftmode{\headrm\quad (\jobname,\ le
\number\day/\number\month/\number\yearltd\ \ \hourmin)}\else{}\fi} 
\newif\iffrancmode
\francmodefalse
% ********* A few math symbols
\def\e{\mathop{\rm e}\nolimits}

\def\d{{\rm d}}
\def\ud{{\textstyle{1\over 2}}}

\def\tr{\mathop{\rm tr}\nolimits}
\def\det{\mathop{\rm det}\nolimits}

\chardef\sigmat=27
\def\rf{\par\item{}}

\def\frac#1#2{{\textstyle{#1\over#2}}}

\def\leaderfill{\leaders\hbox to 1em{\hss.\hss}\hfill}
%%%%%%%%%%%%%%%%%%%%%%%%%%%%%%%%%%%%%%%%%%%%%%%% 
\catcode`\@=11

% ************** double alignment in eqalignno style **********************
\def\deqalignno#1{\displ@y\tabskip\centering \halign to
\displaywidth{\hfil$\displaystyle{##}$\tabskip0pt&$\displaystyle{{}##}$
\hfil\tabskip0pt &\quad
\hfil$\displaystyle{##}$\tabskip0pt&$\displaystyle{{}##}$ 
\hfil\tabskip\centering& \llap{$##$}\tabskip0pt \crcr #1 \crcr}}
%%%%%%%%%%%%%%% double eqalign %%%%%%%%%%%%%%%%%%%%%%%
\def\deqalign#1{\null\,\vcenter{\openup\jot\m@th\ialign{
\strut\hfil$\displaystyle{##}$&$\displaystyle{{}##}$\hfil
&&\quad\strut\hfil$\displaystyle{##}$&$\displaystyle{{}##}$
\hfil\crcr#1\crcr}}\,}
%***************************************************************************
% protection macro for undefined macros
\def\xlabel#1{\expandafter\xl@bel#1}\def\xl@bel#1{#1}
\def\label#1{\l@bel #1{\hbox{}}}
\def\l@bel#1{\ifx\UNd@FiNeD#1\message{label \string#1 is undefined.}%
\xdef#1{?.? }\fi{\let\hyperref=\relax\xdef\next{#1}}%
\ifx\next\em@rk\def\next{}%
%\else\ifx\next#1\xlabel#1\fi\fi\next
\else\def\next{#1}\fi\next}
%***************************************************************************
%********* titlepage, headline, chapter section, subsection, sub, appendix *********
%***************************************************************************
%**************** input with check of file existence ***********************
% Warning macro
\def\DefWarn#1{\ifx\UNd@FiNeD#1\else
\immediate\write16{*** WARNING: the label \string#1 is already defined%
***}\fi}% 
%NOW WORK syntax \cinput{filename} 
\newread\ch@ckfile
\def\cinput#1{\def\filen@me{#1 }% space mandatory after #1 !!
\immediate\openin\ch@ckfile=\filen@me
\ifeof\ch@ckfile\message{<< (\filen@me\ DOES NOT EXIST in this pass)>>}\else% 
\closein \ch@ckfile\input\filen@me\fi}
%********* introduce equation number file: for non-causal quotation
\ifx\UNd@FiNeD\prefix\def\prefix{}\fi % correction added
\newread\ch@ckfile
\immediate\openin\ch@ckfile=\jobname.def
\ifeof\ch@ckfile\message{<< (\jobname.def DOES NOT EXIST in this pass) >>}
\else
\def\DefWarn#1{}%
\closein \ch@ckfile% 
\input\jobname.def\fi
%**********
% Autre utilitaire
\def\listcontent{%\immediate\closeout\tfile%
\immediate\openin\ch@ckfile=\jobname.tab % space mandatory after tab!!
\ifeof\ch@ckfile\message{no file \jobname.tab, no table of contents this
pass}%
\else\closein\ch@ckfile%
\def\sectionname{\iffrancmode Table des
mati\`eres \else Contents\fi}
\centerline{\twbfx\iffrancmode Table des
mati\`eres \else Contents\fi}\nobreak\medskip% 
{\baselineskip=12pt\parskip=0pt\catcode`\@=11\input\jobname.tab
\catcode`\@=12\bigbreak\bigskip}\fi}
%**************************************************************************
\newcount\nochapter
\newcount\nosection
\newcount\nosubsection
\newcount\neqno
\newcount\notenumber
\newcount\nofigure
\newcount\notable
\newcount\noexerc
\newcount\fpage
\newcount\firstpage
\newif\ifappmode
\newwrite\equa
%\newwrite\tab 
%\newwrite\eqdf
% ******************* titlepage **********************************

\newdimen\hulp
\def\maketitle#1{
\edef\oneliner##1{\centerline{##1}}
\edef\twoliner##1{\vbox{\parindent=0pt\leftskip=0pt plus 1fill\rightskip=0pt
plus 1fill 
                     \parfillskip=0pt\relax##1}} 
\setbox0=\vbox{#1}\hulp=0.5\hsize
                 \ifdim\wd0<\hulp\oneliner{#1}\else
                 \twoliner{#1}\fi}
\def\preprint#1{\ifdraftmode\gdef\prepname{\jobname/#1}\else%
\gdef\prepname{#1}\fi\hfill{%\sacfont
\expandafter{\prepname}}\vskip20mm} 
% **************** beginning
\def\title#1\par{\gdef\titlename{#1}
%\global\fpage=\pageno
\global\firstpage=\pageno
\nosection=0
\mark{\the\nosection}
\maketitle{\ssbx\uppercase\expandafter{\titlename}}
\vskip17mm
\nochapter=0
\neqno=0
\notenumber=0
\nofigure=0
\notable=0
\def\prefix{}
\appmodefalse
\mark{\the\nochapter}
\message{#1}
%\immediate\openout\tab=\table
%%%%%%%%%%%%%%%%%%%%%\immediate\openout\equa=\equation%
\immediate\openout\equa=\jobname.equ %
%\immediate\openout\eqdf=\labeldefs
%\edef\ecrire{\write\tab{\par\noindent{\ssbx\ \titlename} 
%\string\leaderfill{\noexpand\number\pageno}}}\ecrire
%\edef\ecrire{\write\equa{{\titlename},
%{\noexpand\number\pageno}, Date \today}\write\equa{}}\ecrire
}
\ifbookmode% 
\headline={\ifnum\pageno=\firstpage\hfill\else\ifodd\pageno\rightheadline
\else\leftheadline\fi\fi}
\else
\headline={\ifnum\pageno=\firstpage\hfill\else\draftdate\hfil{\headrm\folio}%
\hfil\hphantom{\draftdate}\fi}\fi 
%}
\def\abstract{\vskip8mm\iffrancmode\centerline{R\'ESUM\'E}\else%
\centerline{ABSTRACT}\fi \smallskip \begingroup\narrower
\elevenpoint\baselineskip10pt} 

% ***************** input table of contents
%\def\content{\vfill\eject% A un bug dans le format double colonne
%\begingroup\centerline{\uppercase\expandafter{Table of contents}}% 
%\bigskip\elevenpoint\noindent% 
%\parindent=2em
%\openin 1=\jobname.tab
%\ifeof 1\closein1\message{<< (\jobname.tab DOES NOT EXIST. TeX again)>>}%
%\else\input\jobname.tab\closein 1\fi 
%\endgroup\vfill\eject}
%******************************* SECTION ****************************
%******************************************************************
\def\section#1\par{\vskip0pt plus.1\vsize\penalty-100\vskip0pt plus-.1
\vsize\bigskip\vglue\parskip\par%
\global\fpage=\pageno
\ifnum\nochapter=0\ifappmode\relax\else\writetoc
\fi\fi% ajout
\advance\nochapter by 1\global\nosection=0\global\neqno=0
\gdef\sectionname{#1}
%%%%%%%%%%%%%%%%%%%%%%%%%%%%%%%%%%%%%%%%%%%%%%
\vbox{\noindent\twbfx{\hyperdef\hypernoname{section}{\prefix\the\nochapter}%
{\prefix\the\nochapter}\ #1}}%
%%%%%%%%%%%%%%%%%%%%%%%%%%%%%%%%%%%%%%%%%
%\vbox{\noindent\twbfx\ifappmode\iffrancmode{Appendice\ }\else {Appendix\ }\fi
%\else\iffrancmode{Chapitre\ }\else {Chapter\ }\fi\fi%
\message{\prefix\the\nochapter\ \sectionname}%
%{\hyperdef\hypernoname{chapter}{\prefix\the\nochapter}{\prefix\the\nochapter}%
%\vskip11.5mm
%\vbox{\noindent\twbfx{#1}}}}\vskip9mm%
\writetoca{{\string\hyperref{}{section}{\prefix\the\nochapter}%
{\prefix\the\nochapter}} {#1}}%
%\mark{\the\nosection
\bigskip\noindent%
}

% appendix
\def\appendix#1#2\par{\bigbreak\nochapter=0
\appmodetrue
\notenumber=0
\neqno=0
\def\prefix{A}
\mark{\the\nochapter}
\message{APPENDICES}
%\leftline{APPENDICES}
{\hyperdef\hypernoname{section}{\prefix}{ 
\leftline{\uppercase\expandafter{#1}}
\leftline{\uppercase\expandafter{#2}}}}
\noindent\nonfrenchspacing
\writetoca{\string\hyperref{}{section}{\prefix}{Appendices}.\ #1 \ #2}%
}
% **************************** SUBSECTION *************************
\def\subsection#1\par{\vskip0pt plus.05\vsize\penalty-100\vskip0pt
plus-.05\vsize\bigskip\vskip1mm\advance\nosection by 1
\global\nosubsection=0
\def\subsectionname{#1}
\mark{#1}
\message{\the\nochapter.\the\nosection\ #1}
\vbox{\noindent\bf{\hyperdef\hypernoname{section}{\prefix\the\nochapter.%
\the\nosection}{\prefix\the\nochapter.\the\nosection\ #1}}}\vskip1mm%
%\mark{\the\nosection}
\smallskip\noindent% 
\writetoca{{\string\hyperref{}{section}{\prefix\the\nochapter.%
\the\nosection}{\prefix\the\nochapter.\the\nosection}} {#1}}%
} 
%%%%%%%%%%%%%%%%%% SUBSUBSECTION %%%%%%%%%%%%%%%%%%%%
\def\ssubsection#1\par{\vskip0pt plus.05\vsize\penalty-100\vskip0pt%
plus-.05\vsize\medskip\vskip\parskip\advance\nosubsection by 1%
\message{\the\nochapter.\the\nosection.\the\nosubsection\ #1}%
\vbox{\noindent\bf{\hyperdef\hypernoname{section}{\prefix\the\nochapter.%
\the\nosection.\the\nosubsection}{\prefix\the\nochapter.\the\nosection.%
\the\nosubsection\ \bf #1}}}%
%\vbox{\noindent\it\prefix\the\nochapter.\the\nosection.\the\nosubsection\
%\it #1}
\smallskip\noindent% 
\writetoca{{\string\hyperref{}{section}{\prefix\the\nochapter.%
\the\nosection.\the\nosubsection}{\prefix\the\nochapter.\the\nosection.%
\the\nosubsection}} {#1}}%
}   
%%%%%%%%%%%%%%%%%%%%%%%%%%%%%%%%%%%%%%%%%%%%%%%
%
\def\note #1{\advance\notenumber by 1
\footnote{$^{\the\notenumber}$}{\sevenrm #1}} 
% ?????

%\def\enchapter{\end}
\parindent=1em 
\newinsert\margin
\dimen\margin=\maxdimen
\count\margin=0 \skip\margin=0pt
%*****************************************************************
% IMPORTANT, new version demands chapter be defined before any section,
% section be defined before any subsection
\def\sslbl#1{\DefWarn#1%
\ifdraftmode{\leavevmode\vadjust{\smash%
{\line{{\escapechar=` \hfill\rlap{\sevenrm\hskip1mm\string#1}}}}}}%
\fi 
\ifnum\nochapter=0%
\if\prefix{}\xdef#1{}%
\edef\ewrite{\write\equa{{\string#1}}%
\write\equa{}}\ewrite%
\else
%%%%%%%%%%%%%%%%%%%%%%%%%%%%%%%%%%%%%%
\xdef#1{\noexpand\hyperref{}{section}{\prefix}{\prefix}}%
\edef\ewrite{\write\equa{{\string#1},\prefix}%
\write\equa{}}\ewrite%
\writedef{#1\leftbracket#1}
%%%%%%%%%%%%%%%%%%%%%%%%%%%%%%%%%%%%%%
\fi
\else%
\ifnum\nosection=0%
\xdef#1{\noexpand\hyperref{}{section}{\prefix\the\nochapter}%
{\prefix\the\nochapter}}%
\edef\ewrite{\write\equa{{\string#1},\prefix\the\nochapter}%
\write\equa{}}\ewrite%
\writedef{#1\leftbracket#1}
\else%
\ifnum\nosubsection=0%
\xdef#1{\noexpand\hyperref{}{section}{\prefix\the\nochapter.%
\the\nosection}{\prefix\the\nochapter.\the\nosection}}%
\writedef{#1\leftbracket#1}
\edef\ewrite{\write\equa{{\string#1},\prefix\the\nochapter%
.\the\nosection}\write\equa{}}\ewrite%
\else%
\xdef#1{\noexpand\hyperref{}{section}%
{\prefix\the\nochapter.\the\nosection.\the\nosubsection}%
{\prefix\the\nochapter.\the\nosection.\the\nosubsection}}%
\writedef{#1\leftbracket#1}
%\xdef#1{\prefix\the\nochapter.\the\nosection.\the\nosubsection}%
%\edef\ewrite{\write\eqdf{\string\def\string#1{\prefix\the\nochapter.%
%\the\nosection.\the\nosubsection}}\write\eqdf{}}\ewrite%
\edef\ewrite{\write\equa{{\string#1},\prefix\the\nochapter.\the\nosection%
.\the\nosubsection}\write\equa{}}\ewrite%
\fi\fi\fi}%
%**********************************************************************

% *************
\newwrite\tfile \def\writetoca#1{}
%       use this to write file with table of contents
\def\writetoc{\immediate\openout\tfile=\jobname.tab
\def\writetoca##1{{\edef\next{\write\tfile{\noindent ##1 \string\leaderfill%
%{\string\hyperref{}{page}{\noexpand\number\pageno}{\noexpand\number\pageno}}
\noexpand\number\pageno\par}}\next}}}

% ********************* references harvmac style ***********************
%     \ref\label{text}
% generates a number, assigns it to \label, generates an entry.
% To list the refs on a separate page,  \listrefs
%
\def\nolabels{\def\wrlabeL##1{}\def\eqlabeL##1{}\def\reflabeL##1{}}
\def\writelabels{\def\wrlabeL##1{\leavevmode\vadjust{\rlap{\smash%
{\line{{\escapechar=` \hfill\rlap{\sevenrm\hskip.03in\string##1}}}}}}}%
\def\eqlabeL##1{{\escapechar-1\rlap{\sevenrm\hskip.05in\string##1}}}%
\def\reflabeL##1{\noexpand\llap{\noexpand\sevenrm\string\string\string##1}}}
\ifdraftmode\writelabels\else\nolabels\fi

\global\newcount\refno \global\refno=1
\newwrite\rfile
\def\ref{[\hyperref{}{reference}{\the\refno}{\the\refno}]\nref}
\def\nref#1{\DefWarn#1%
\xdef#1{[\noexpand\hyperref{}{reference}{\the\refno}{\the\refno}]}%
\writedef{#1\leftbracket#1}%
\ifnum\refno=1\immediate\openout\rfile=\jobname.ref\fi
\chardef\wfile=\rfile\immediate\write\rfile{\noexpand\item{[\noexpand\hyperdef%
\noexpand\hypernoname{reference}{\the\refno}{\the\refno}]\ }%
\reflabeL{#1\hskip.31in}\pctsign}\global\advance\refno by1\findarg}
%       horrible hack to sidestep tex \write limitation%
\def\findarg#1#{\begingroup\obeylines\newlinechar=`\^^M\pass@rg}
{\obeylines\gdef\pass@rg#1{\writ@line\relax #1^^M\hbox{}^^M}%
\gdef\writ@line#1^^M{\expandafter\toks0\expandafter{\striprel@x #1}%
\edef\next{\the\toks0}\ifx\next\em@rk\let\next=\endgroup\else\ifx\next\empty%
\else\immediate\write\wfile{\the\toks0}\fi\let\next=\writ@line\fi\next\relax}}
\def\striprel@x#1{} \def\em@rk{\hbox{}}
\def\lref{\begingroup\obeylines\lr@f}
\def\lr@f#1#2{\DefWarn#1\gdef#1{\let#1=\UNd@FiNeD\ref#1{#2}}\endgroup\unskip}
\def\semi{;\hfil\break}
\def\addref#1{\immediate\write\rfile{\noexpand\item{}#1}} %now unnecessary
\def\listrefs{{}\vfill\supereject\immediate\closeout\rfile\writestoppt
\baselineskip=14pt
\gdef\reference{\iffrancmode  R\'eferences \else References\fi}
\mark{\reference}  
\nochapter=0\def\sectionname{\reference} 
%%%%%%%%%%%%%%%%%%%%%%%%%%%%%%%%%
{\centerline{\bf \hyperdef\hypernoname{refer}{refer}{\bf \reference}}}
\nobreak\bigskip\noindent 
\writetoca{\string\hyperref{}{refer}{refer}{\reference}}
%%%%%%%%%%%%%%%%%%%%%%%%%%%%%%%%%
{\parindent=20pt%
\frenchspacing\escapechar=` \input \jobname.ref\vfill\eject}\nonfrenchspacing}
\def\startrefs#1{\immediate\openout\rfile=\jobname.ref\refno=#1}
\def\xref{\expandafter\xr@f}\def\xr@f[#1]{#1}
\def\refs#1{\count255=1[\r@fs #1{\hbox{}}]}
\def\r@fs#1{\ifx\UNd@FiNeD#1\message{reflabel \string#1 is undefined.}%
\nref#1{need to supply reference \string#1.}\fi%
\vphantom{\hphantom{#1}}{\let\hyperref=\relax\xdef\next{#1}}%
\ifx\next\em@rk\def\next{}%
\else\ifx\next#1\ifodd\count255\relax\xref#1\count255=0\fi%
\else#1\count255=1\fi\let\next=\r@fs\fi\next}
%************************
%*******
%
\newwrite\lfile
{\escapechar-1\xdef\pctsign{\string\%}\xdef\leftbracket{\string\{}
\xdef\rightbracket{\string\}}\xdef\numbersign{\string\#}}
\def\writedefs{\immediate\openout\lfile=\jobname.def \def\writedef##1{%
{\let\hyperref=\relax\let\hyperdef=\relax\let\hypernoname=\relax
 \immediate\write\lfile{\string\def\string##1\rightbracket}}}}%
\def\writestop{\def\writestoppt{\immediate\write\lfile{\string\pageno%
\the\pageno\string\startrefs\leftbracket\the\refno\rightbracket%
\string\def\string\secsym\leftbracket\secsym\rightbracket%
\string\secno\the\secno\string\meqno\the\meqno}\immediate\closeout\lfile}}
\def\writestoppt{}\def\writedef#1{}
\writedefs
% ******
% bibliography: not very satisfactory
\def\biblio\par{\vskip0pt plus.1\vsize\penalty-100\vskip0pt plus-.1
\vsize\bigskip\vskip\parskip
\message{Bibliographie}
{\leftline{\bf \hyperdef\hypernoname{biblio}{bib}{Bibliographical Notes}}}
\nobreak\medskip\noindent\frenchspacing
\writetoca{\string\hyperref{}{biblio}{bib}{Bibliographical Notes}}}%

%**************** autre version si plusieurs biblio ************************
\def\biblionote{\iffrancmode Notes Bibliographiques\else Bibliographical Notes
\fi}
\def\beginbib\par{\vskip0pt plus.1\vsize\penalty-100\vskip0pt plus-.1
\vsize\bigskip\vskip\parskip
\message{Bibliographie}
{\leftline{\bf \hyperdef\hypernoname{biblio}{\prefix\the\nochapter}%
{\biblionote}}}
\nobreak\medskip\noindent\frenchspacing
\writetoca{\string\hyperref{}{biblio}{\prefix\the\nochapter}%
{\biblionote}}}%

% *************** exercises: same comment
\def\Exercises{\iffrancmode Exercices\else Exercises
\fi}
\def\exerc\par{\vskip0pt plus.1\vsize\penalty-100\vskip0pt plus-.1
\vsize\bigskip\vskip\parskip\global\noexerc=0
\message{Exercises}
\iffrancmode\mark{Exercices}\else\mark{Exercises}\fi
{\leftline{\bf\hyperdef\hypernoname{exercise}{\the\nochapter}{\Exercises}}}
\nobreak\medskip\noindent\frenchspacing
\writetoca{\string\hyperref{}{exercise}{\the\nochapter}{\Exercises}}
}
\def\esubsec{\ifnum\noexerc=0\vskip-12pt\else\vskip0pt plus.05\vsize%
\penalty-100\vskip0pt plus-.05\vsize\bigskip\vskip\parskip\fi%
\global\advance\noexerc by 1
\hyperdef\hypernoname{exercise}{\the\nochapter.\the\noexerc}{}%
\vbox{\noindent\it \iffrancmode Exercice\else Exercise\fi\ \the\nochapter.\the\noexerc}\smallskip\noindent}
%%%%%%%%%%%%%%%%%%%%%%%%%%%
\def\exelbl#1{\ifdraftmode{\hfill\escapechar-1{\rlap{\hskip-1mm%
\sevenrm\string#1}}}\fi%
{\xdef#1{\noexpand\hyperref{}{exercise}{\the\nochapter.\the\noexerc}%
{\the\nochapter.\the\noexerc}}}%
\edef\ewrite{\write\equa{{\string#1}\the\nochapter.\the\noexerc}%
\write\equa{}}\ewrite%
\writedef{#1\leftbracket#1}}
%****************************

%*************************************************************************
%Macro de numerotation automatique
%*************************************************************************
% numbering without naming
\def\eqnn{\global\advance\neqno by 1 \ifinner\relax\else%
\eqno\fi(\prefix\the\nochapter.\the\neqno)}
%
% numbering and attaching a name: \eqnd{\ename}
\def\eqnd#1{\DefWarn#1%
\global\advance\neqno by 1 
{\xdef#1{($\noexpand\hyperref{}{equation}{\prefix\the\nochapter.\the\neqno}%
{\prefix\the\nochapter.\the\neqno}$)}}%???
\ifinner\relax\else\eqno\fi(\hyperdef\hypernoname{equation}{\prefix\the%
\nochapter.\the\neqno}{\prefix\the\nochapter.\the\neqno})
\writedef{#1\leftbracket#1}
\ifdraftmode{\escapechar-1{\rlap{\hskip.2mm\sevenrm\string#1}}}\fi
\edef\ewrite{\write\equa{{\string#1},(\prefix\the\nochapter.\the\neqno)
{\noexpand\number\pageno}}\write\equa{}}\ewrite}
%
% for eqalignno, allows (1a) (1b)...
\def\checkm@de#1#2{\ifmmode{\def\f@rst##1{##1}\hyperdef\hypernoname{equation}%
{#1}{#2}}\else\hyperref{}{equation}{#1}{#2}\fi}
\def\f@rst#1{\c@t#1a\em@ark}\def\c@t#1#2\em@ark{#1}
\def\eqna#1{\DefWarn#1%
\global\advance\neqno by1\ifdraftmode{\hfill%
\escapechar-1{\rlap{\sevenrm\string#1}}}\fi%
\xdef #1##1{(\noexpand\relax\noexpand%
\checkm@de{\prefix\the\nochapter.\the\neqno\noexpand\f@rst{##1}1}%
{\hbox{$\prefix\the\nochapter.\the\neqno##1$}})}
\writedef{#1\numbersign1\leftbracket#1{\numbersign1}}%
} 
\def\em@rk{\hbox{}} 
\def\xeqn{\expandafter\xe@n}\def\xe@n(#1){#1}
\def\xeqna#1{\expandafter\xe@na#1}\def\xe@na\hbox#1{\xe@nap #1}
\def\xe@nap$(#1)${\hbox{$#1$}}
% \eqns allows to quote several equations, suppressing unnecessary ()
\def\eqns#1{(\e@ns #1{\hbox{}})}
\def\e@ns#1{\ifx\UNd@FiNeD#1\message{eqnlabel \string#1 is undefined.}%
\xdef#1{(?.?)}\fi{\let\hyperref=\relax\xdef\next{#1}}%
\ifx\next\em@rk\def\next{}%
\else\ifx\next#1\xeqn#1\else\def\n@xt{#1}\ifx\n@xt\next#1\else\xeqna#1\fi
\fi\let\next=\e@ns\fi\next}
%**********************************************************************
%*************************** figure macros ****************************
% Pour centrer ajouter 16mm a la taille de la boite
\def\figure#1#2{\global\advance\nofigure by 1 \vglue#1%
\hyperdef\hypernoname{figure}{\the\nofigure}{}%
{\elevenpoint
\setbox1=\hbox{#2}
\ifdim\wd1=0pt\centerline{Fig.\ \the\nofigure\hskip0.5mm}%
\else\def\caption{Fig.\ \the\nofigure\quad#2\hskip0mm}
\setbox0=\hbox{\caption}
\ifdim\wd0>\hsize\noindent Fig.\ \the\nofigure\quad#2\else
                 \centerline{\caption}\fi\fi}\par}
% le bigskip a la fin a ete enleve!
 % obsolete, for compatibility
%***************
%figure alignee a gauche
\def\lfigure#1#2{\global\advance\nofigure by
1\vglue#1%
\hyperdef\hypernoname{figure}{\the\nofigure}{}%
\leftline{\elevenpoint\hskip10truemm  Fig.\
\the\nofigure\quad #2}} 
%***************
\def\figlbl#1{\ifdraftmode{\hfill\escapechar-1{\rlap{\hskip-1mm%
\sevenrm\string#1}}}\fi%
{\xdef#1{\noexpand\hyperref{}{figure}{\the\nofigure}%
{\the\nofigure}}}%
\edef\ewrite{\write\equa{{\string#1}\the\nofigure}%
\write\equa{}}\ewrite%
\writedef{#1\leftbracket#1}}
%****************************
\def\tablbl#1{\global\advance\notable by
1\ifdraftmode{\hfill\escapechar-1{\rlap{\hskip-1mm%
\sevenrm\string#1}}}\fi%
\hyperdef\hypernoname{table}{\the\notable}{}
{\xdef#1{\noexpand\hyperref{}{table}{\the\notable}%
{\the\notable}}}%
\edef\ewrite{\write\equa{{\string#1}\the\notable}%
\write\equa{}}\ewrite%
\writedef{#1\leftbracket#1}}

%***********************************************************************
%***********************************************************************
\catcode`@=12

\catcode`\À=\active\defÀ{\ifmmode\grave A\else\`A\fi}
\catcode`\É=\active\defÉ{\ifmmode\acute E\else\'E\fi}
\catcode`\à=\active\defà{\ifmmode\grave a\else\`a\fi}
\catcode`\â=\active\defâ{\ifmmode\hat a\else\^a\fi}
\catcode`\ä=\active\defä{\ifmmode\ddot a\else\"a\fi}
\catcode`\ç=\active\defç{\c c}
\catcode`\é=\active\defé{\ifmmode\acute e\else\'e\fi}
\catcode`\è=\active\defè{\ifmmode\grave e\else\`e\fi}
\catcode`\ê=\active\defê{\ifmmode\hat e\else\^e\fi}
\catcode`\ë=\active\defë{\ifmmode\ddot e\else\"e\fi}
\catcode`\ï=\active\defï{\ifmmode\ddot i\else\"\i\fi}
\catcode`\î=\active\defî{\ifmmode\hat i\else\^\i\fi}
\catcode`\ö=\active\defö{\ifmmode\ddot o\else\"o\fi}
\catcode`\ô=\active\defô{\ifmmode\hat o\else\^o\fi}
\catcode`\ù=\active\defù{\ifmmode\grave u\else\`u\fi}
\catcode`\ü=\active\defü{\ifmmode\ddot u\else\"u\fi}
\catcode`\û=\active\defû{\ifmmode\hat u\else\^u\fi}

\input epsf
%\draft
\title{Random Vector and Matrix Theories: A Renormalization Group Approach} 

\centerline{{\bf JEAN ZINN-JUSTIN}} 
\bigskip
{\baselineskip14pt\centerline{CEA/IRFU} 
\smallskip
\centerline{Centre de Saclay,}
\centerline{91191 Gif-sur-Yvette Cedex, France}}
%\centerline{\Blue{E-mail:jean.zinn-justin@.cea.fr}}}
\vskip8mm

\footnote{}{\it In memoriam of K.G.~Wilson}
\footnote{}{\it Journal of Statistical Physics: to be published} 
\vfill\eject 
The study of the statistical properties of {\it random matrices of large size}\/ has a long history, dating back to Wigner, who suggested using {\it Gaussian ensembles}\/ to give a statistical description of the spectrum of complex Hamiltonians and  derived the famous semi-circle law, the work of Dyson and many others. Later, 't Hooft noticed that, in  $SU(N)$  non-Abelian gauge theories, {\it tessalated surfaces  can be associated to Feynman diagrams}\/ and that the   large  $N$  expansion corresponds to an expansion in successive topologies. \par
Following this observation, some times later, it was realized that some ensembles of random matrices in the   {\it large  $N$  expansion}\/ and the so-called {\it double scaling limit}\/ could be used as toy models for quantum gravity: {\it 2D quantum gravity  coupled to conformal matter}. This has resulted in a tremendous expansion of  random matrix theory, tackled with increasingly sophisticated mathematical methods and number of matrix models have been solved exactly. However, the somewhat paradoxical situation is that either models can be solved exactly or little can be said. 
\par
 Since the solved models display {\it critical points}\/ and {\it universal properties}, it is tempting to use   renormalization group  (RG) ideas to determine universal properties, without solving models explicitly.  Initiated by Brézin and Zinn-Justin, the approach  has led to encouraging results, first for matrix integrals and then quantum mechanics with matrices, but has not yet become a universal tool as initially hoped. In particular, general quantum field theories with matrix fields require more detailed investigations. \par
To better understand some of the encountered difficulties, we first apply analogous ideas to the simpler  $O(N)$  symmetric vector models, models that  can  be solved quite generally in the large  $N$  limit. Unlike other attempts, our method is a close extension of Brézin and Zinn-Justin. Discussing vector and matrix models with similar approximation scheme,   we notice that in all cases (vector and matrix integrals,   vector and matrix path integrals in the local approximation), at leading order, non-trivial fixed points satisfy the same {\it universal algebraic equation}, and this is the main result of this work. However, its precise meaning and role have still to be better understood. 
\smallskip
\nref\rMMZJ{M. Moshe and  J. Zinn-Justin, {\it Quantum field theory in the large N limit: a review}, {\bf  Phys. Rep.} 385 (2003) 69-228.}
\nref\rONscaling{P. Di Vecchia, M. Kato, N. Ohta, {\it Double scaling limit in O(N) vector models}, {\bf Nucl. Phys.} B357 (1991) 495-520; A. Anderson, R.C. Myers, V.Periwal, {\it Branched polymers from a double-scaling limit of matrix models}, {\bf Nucl. Phys.} B360 (1991) 463-479.}  \nref\rZJONscaling{J. Zinn-Justin, {\it O(N) vector field theories in the double
scaling limit}, {\bf Phys. Lett.} B257 (1991) 335-430.}
\nref\rEBZJ{E. Brézin and J. Zinn-Justin, {\it Renormalization group
approach to matrix models}, {\bf Phys. Lett.} B288 (1992) 54-58.}
\nref\rKGWJK{K.G. Wilson and J. Kogut, {\it The renormalization group and the $\epsilon$ expansion},  {\bf Phys. Rep.} 12C (1974) 75-199.}
\nref\rSHCINS{S. Higuchi, C. Itoi, N Sakai, {\it Nonlinear renormalization group equation for matrix models}, {\bf Phys. Lett.} B318 (1993) 63-72; {\it Exact beta functions in the vector model and renormalization group approach},   {\bf Phys. Lett.} B312 (1993) 88-96.}
\nref\rDKAJ{F. David, {\it Planar diagrams, two-dimensional lattice gravity and surface models}, {\bf Nucl. Phys.} B257 (1985) 45-58; {\it A model of random surfaces with non-trivial critical behaviour}, {\bf Nucl. Phys.\/} B257 (1985) 543-576\semi
V.A. Kazakov, {\it Bilocal regularization of models of random surfaces},\goodbreak {\bf Phys. Lett.} {\bf B150} (1985) 282-284;\rf
J. Ambj\o rn, B. Durhuus and J. Fröhlich, {\it Diseases of triangulated random surface models, and possible cures}, {\bf Nucl. Phys.} {\bf B257}
(1985) 433-449.}
\nref\rBKDSGM{E. Brézin and V.A. Kazakov, {\it Exactly solvable field theories of closed strings}, {\bf Phys. Lett.\/} {\bf B236} (1990) 144-150\semi
M.R. Douglas and S.H. Shenker, {\it Strings in less than one dimension},  {\bf Nucl. Phys.\/} {\bf B235} (1990) 635-654\semi
D.G. Gross and A.A. Migdal, {\it Nonperturbative two-dimensional quantum gravity}, {\bf Phys. Rev. Lett.\/} {\bf 64} (1990) 127-130.}
\nref\rBKZGP{
E. Brézin, V.A. Kazakov and Al. A. Zamolodchikov, {\it Scaling violation in a field theory of closed strings in one physical dimension}, {\bf Nucl. Phys.\/}
{\bf B338} (1990) 673-688\semi
P. Ginsparg and J. Zinn-Justin, {\it 2d gravity + 1d matter}, {\bf Phys. Lett.} {\bf B240} (1990) 333-340\semi
D.J. Gross and N. Miljkovic, {\it A nonperturbative solution of $D = 1$ string theory}, {\bf Phys. Lett.\/} {\bf B238} (1990) 217-223\semi
G. Parisi, {\it On the one dimensional discretized string}, {\bf Phys. Lett.\/} {\bf B238} (1990) 209-212.} 
\nref\rPDFPGZJ{For a review on early developments in random matrix theory and 2D quantum gravity see {\it e.g.},  
Ph.~Di Francesco, P.~Ginsparg and J.~Zinn-Justin, {\it 2D
gravity and random matrices}, {\bf Phys.~Rept.}  254 (1995)  1-133.}
\nref\rmatrg{S. Higuchi, C. Itoi, S. Nishigaki and N. Sakai, {\it Nonlinear renormalization group equation for matrix models}, {\bf Phys. Lett.} B318 (1993) 63-72; {\it Renormalization group flow in one- and two-matrix models}, {\bf Nucl. Phys.}
B434 (1995) 283-318; {\it Renormalization group approach to multiple-arc random matrix models}, {\bf Phys. Lett.} B398 (1997) 123-129\semi G. Bonnet, F. David, {\it Renormalization group for matrix models with branching interactions}, {\bf Nucl. Phys.} B552 (1999) 511-528.}

\nref\rJAPHD{J. Alfaro and   P.H. Damgaard, {\it The $D=1$ matrix model and the renormalization group}, {\bf Phys. Lett.} B289 (1992) 342-346.}
\nref\rSDTD{S. Dasgupta, T Dasgupta, {\it Renormalization Group Approach to c=1 Matrix Model on a circle and D-brane Decay}, {\bf arXiv preprint} hep-th/0310106.}
\nref\rSNishi{S. Nishigaki, {\it Wilsonian approximated renormalization group for matrix and vector models in $2 < d < 4$}, {\bf Phys. Lett.} B376 (1996) 73-81.}
\vfill\eject
 
\section{$O(N)$ symmetric vector models: simple integrals}

$O(N)$ symmetric vector models can be solved, quite generally, in the large $N$ limit \refs{\rMMZJ}. The large $N$ behaviour and critical properties of  $O(N)$-symmetric {\it vector models}\/ can then be studied in the same spirit as matrix models but the analysis can be extended also to {\it path and field integrals}\/ and analogous {\it double scaling limits}\/ can then be exhibited \refs{\rONscaling, \rZJONscaling}. \par 
In the case of simple integrals, the geometric interpretation of critical points is related to the statistical properties of a class of continuous chains  (obtained as limits of chains of bubble diagrams, see figure \label{\figtwentynine}), of the form of {\it branched polymers or filamentary surfaces}, classified by the amount  of stretching and the number of vertices (see figure \label{\figrtnine}).\par
\midinsert
\epsfxsize=90mm
\epsfysize=10mm
\centerline{\epsfbox{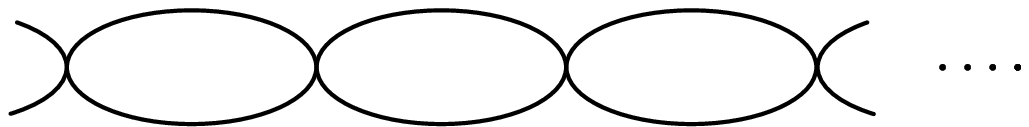}}
\figure{2.mm}{Vector models: The dominant diagrams in the large $N$ limit.}
\endinsert
\figlbl{\figtwentynine}
\midinsert
\epsfxsize=17.4mm
\epsfysize=17.3mm
\centerline{\epsfbox{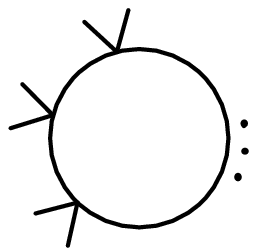}}
\figure{2.mm}{Vector models: Vertices in the large $N$ limit.}
\endinsert
\figlbl{\figrtnine}
Here, we first recall how various $O(N)$ symmetric vector integrals can be calculated in the large $N$ limit and exhibit  {\it multicritical points and scaling relations}. We then try to recover the results by RG methods extending those used for matrix models \refs{\rEBZJ}, in order to better understand the difficulties encountered in the latter case.
\subsection{Zero dimension or simple integrals}

We   consider the integral 
$$\e^{Z_N }=\left(N\over2\pi\right)^{N/2}\int\d^N {\bf x}\,\e^{-NV({\bf x}^2)},\eqnd\eIntegral$$ 
where  ${\bf x}$  is an $N$-component vector and we first choose the normalization  $V(\rho)=\rho/2+O(\rho^2)$. We assume that the function $V(\rho)$  is analytic in a neighbourhood of the real positive axis, the simplest example being a polynomial.\par
To evaluate the integral, we first integrate over angles, which reduces the integral to ($\rho={\bf x}^2$)
$$\e^{Z_N }={\cal N}\int{\d \rho\over \rho}\e^{-N\sigma(\rho)} \eqnd\eIntegralb $$ 
 with
$$\sigma(\rho)=V(\rho)-\ud\ln \rho $$ 
and
$${\cal N}=\left(N\over2\pi\right)^{N/2}{ \pi^{N/2}\over\Gamma(N/2)}\mathop{\sim}_{N\to\infty} \sqrt{N\over 4\pi   }\e^{N/2}\, .$$ 
The large $N$ limit can be determined by using the steepest descent method. The saddle point equation $\rho_c$  is given by
$$\sigma'(\rho_c)=0\ \Leftrightarrow\ 2V'(\rho_c)\rho_c=1\,. \eqnd\esadd$$ 
Expanding to  $\sigma(\rho)$   up order  $(\rho-\rho_c)^2$  and performing the Gaussian integration, one finds
$$ Z_N  =N\left[\ud -V(\rho_c)+\ud \ln\rho_c \right]-\ud\ln\left[2\rho_c^2V''(\rho_c)+1 \right]+O(1/N). $$ 
For later purpose, it is convenient to introduce the function 
$$R(\rho)=1/2V'(\rho) .\eqnd\eIntUdef $$  
The saddle point equation then reduces to
$$R(\rho)  =\rho\,.\eqnd\eIntUsaddle  $$  

\subsection Critical and multicritical points

Critical points \refs{\rONscaling, \rZJONscaling} correspond to situations where a number of derivatives of  $2V'(\rho)-1/\rho$  vanish at the saddle point  $\rho=\rho_c$.\sslbl\ssImulticritical  
At a (multi)critical point, 
$$\sigma(\rho)-\sigma(\rho_c) \propto (\rho-\rho_c)^m, \ m \ge 2\,,  $$ 
and relevant values of  $\rho-\rho_c$  are of order  $N^{-1/m}$, which leads to non-trivial scaling properties. For $m>2$, the fluctuations at the saddle point are no longer  Gaussian. \par
Note that for $m$  odd, the saddle point is only relevant if the integral is defined as a contour integral after some appropriate analytic continuation.  A similar situation is found in matrix models. The problem will arise again in quantum  mechanics and field theory and thus an
analytic continuation will be, when necessary,  implicitly assumed.\par
We normalize  $\rho$  such that the saddle point is located at  $\rho=m-1$,  $m$  integer with  $m\ge 2$,
and assume that the critical potential
$V_c$  satisfies
$$  2V'_c(\rho)\rho-1= -\left(1-{\rho\over m-1}\right)^{m-1} +O\left((\rho-m+1)^m\right) \,.$$ 
Then,
$$\sigma_c(\rho)\equiv V_c(\rho)-\ud\ln\rho= {1\over 2m}\left(1-{\rho\over m-1}\right)^m+O\left((\rho-m+1)^{m+1}\right).$$ 
Note that a strict equality is impossible since  $V(\rho)$  is regular at  $\rho=0$.
Returning to the integral and setting  $z=1-\rho/(m-1)$,
we obtain
$$\e^{Z_N} \propto\int \d z\,\exp\left[-N {z^m\over2m}+O(z^{m+1} ) \right]. $$ 
We see that the values of  $z=1-\rho/(m-1)$  contributing to the
integral are of order  $N^{1/m}$. 
A general relevant (in the RG sense) perturbation to the 
critical function  $V_c$ has then the form
 $$V_q(\rho)= v_q z^q +O(z^{q+1}) ,\ v_q\ne 0\,,\ 1<q<m-1\,, \eqnd\eImulticritical $$ 
($q=m-1$  corresponds to a translation of  $z$) and  the  partition function  $Z_N$  is given for $N$ large and $v_q$  small
by  
 $$\e^{Z_N} \propto\int \d z\,\exp\left[-N{ z^m\over2m}+N\sum_{q=1}^{m-2}v_q
z^q \right].\eqnd\elogZ$$ 
Rescaling  $z$  into  $z N^{-1/m}$  we see that the scaling region
corresponds to take   $v_q$  vanishing with $N$ like  $N^{q/m-1}$.\par
For  $q>m$, the perturbations are {\it irrelevant}\/ and, at  $v_q$  fixed, their contributions vanish like  $N^{q/m-1}$. For $m=2$, that is, for the Gaussian integral, all perturbations are irrelevant.

\section{A renormalization group (RG) inspired strategy}

Studying some specific properties of $O(N)$ vector models in the large $N$ limit, we have exhibited critical points and scaling laws. We are thus reminded of the general theory of critical phenomena \refs{\rKGWJK}, where critical properties can be derived using RG arguments without solving models explicitly. Even though in the large $N$ limit, most vector models  can be solved  exactly, having in mind an application to {\it matrix models}, we try to recover the vector model  results by simple-minded renormalization group considerations.\sslbl\ssIntRG \par
We  expect that a change $ N \mapsto N+\delta N $ can
be compensated by a change of the parameters of the model with the same
continuum physics. We show here that this is the case, at least in some approximation scheme, at the expense of enlarging the space of coupling constants very much as in the Wilson's scheme 
of integration over the momenta in the 
shell $ \Lambda - \d  \Lambda  < |p| < \Lambda  $  and, which in the Wegner--Wilson form, leads to functional RG equations \refs{\rKGWJK}.
Some aspects of this problem for simple integrals have been investigated previously, but in a different spirit in \refs{\rSHCINS}. Here, we are only interested in   methods that can  readily be generalized to path integrals and corresponding matrix integrals \refs{\rEBZJ}, and this requires some approximation scheme.\par

For this purpose, we consider the $O(N+1)$ integral and integrate over only one variable:
$$
\e^{Z_{N+1} }=\left(N+1\over2\pi\right)^{(N+1)/2}\int\d^N {\bf x}\int \d y\,\e^{-(N+1)V({\bf x}^2+y^2)}.$$ 
For  $N\to\infty$, we can expand in powers of  $y$. Keeping the leading term and integrating, we find
 $$\e^{Z_{N+1} }=\left(N+1\over2\pi\right)^{N/2}  \int\d^N {\bf x}\,\e^{-(N+1)V({\bf x}^2)-\ud\ln 2V'({\bf x}^2)}\left(1+O(1/N)\right),$$  
the leading relative correction  being
 $$-{3\over 8N}{V''({\bf x}^2)\over V'{}^2({\bf x}^2)}.$$ 
This confirms that for this method to work  $V'(\rho)\propto R^{-1}(\rho)$  must be a strictly positive function,  a condition that we assume from now on.\par
In the spirit of the RG, we  then  renormalize the integration vector,
 $${\bf x}\mapsto {\bf x}\left(1-{1\over 2N}(1+  \gamma)\right),$$ 
where  $ \gamma$  is a free parameter. We infer,
 $$ Z_{N+1}(V)  = -\frac{1}{2} \gamma+Z_N( V+\delta V)+O(N^{-1}) $$ 
with  
 $$N\delta V(\rho)= V(\rho)- (1+   \gamma)\rho V'(\rho)+\ud \ln 2V'(\rho)+O(N^{-1}).\eqnd\eIRGgen $$ 
\par
\subsection General flow equation

From the variation of the function $V(\rho)$, we infer, for $N\to\infty$,
$$\eqalignno{&N[Z_{N+1}(V)-Z_N(V)]\sim N{\partial \over \partial N}Z_N(V)=-N\gamma \cr&\quad +  \int\d\rho\left[ V(\rho)- (1+  \gamma )\rho V'(\rho)+\ud \ln 2V'(\rho)\right]{\delta\over \delta V(\rho)}Z_N(V),&\eqnn \cr}$$
where for $N$  large  we have treated $1/N$ as a continuous parameter and $\gamma$ is chosen such that $$V(\rho)=\ud\rho+O(\rho^2)\ \Rightarrow\ \gamma=2V''(0).$$  
Correspondingly, after a rescaling $N\mapsto \lambda N$, we find for $V$ the  general flow equation
$$\lambda{\d \over \d\lambda}V(\rho,\lambda)=V(\rho,\lambda)- \bigl(1+  \gamma (\lambda)\bigr)\rho V'(\rho,\lambda)+\ud \ln 2V'(\rho,\lambda),\eqnd\eIRGflowgen $$
where $N\to\infty$ corresponds to $\lambda\to\infty$.
It is convenient to differentiate  the flow equation with respect to $\rho$. In terms of the function
$$R(\rho,\lambda)=1/2V'(\rho,\lambda),\eqnd\eRONzerodef $$
one obtains 
$$\lambda{\d \over \d\lambda}R(\rho,\lambda)=\gamma(\lambda)R(\rho,\lambda)- \bigl(1+  \gamma(\lambda)\bigr) \rho R'(\rho,\lambda)+  R'(\rho,\lambda) R(\rho,\lambda)   \eqnd\eIRGUflowgen $$
with $R(0,\lambda)=1$ and, thus, $\gamma(\lambda)=-R'(0,\lambda)$. The  equation that will again appear in all more general problems.

\section {Fixed points}

To equation \eqns{\eIRGflowgen}, corresponds   the fixed point equation  
$$V(\rho)-(1+\gamma)\rho V'(\rho)+\ud\ln 2V'(\rho)=0\,.$$
We first discuss the equation in a perturbative and then linear approximation.
We then consider the full equation but conveniently expressed in terms of $R=1/2V'$.

\subsection{Critical point: Perturbative approximation}

We first consider the simple example
$$V(\rho)=\ud\rho+\frac{1}{4}g\rho^2.$$
Since the contribution $\ud\ln 2V'(\rho)$ is no longer a polynomial, we assume that we can use a small $ \rho$ approximation   and expand up to order $ \rho^2$:
$$\ud\ln 2V'(\rho)=g\rho-\ud g^2\rho^2+O( \rho^3).$$
The variation of $V$ then reduces to a variation  of $g$. Taking into account that $\gamma=g$, one finds
$$\lambda{\partial g \over\partial \lambda}=-\beta(g)$$
with
$$\beta(g)=g\left(1+3g\right) .$$
This equation determines two fixed points $g^*=0$ with $\beta'(g^*)=1$, the attractive Gaussian fixed point, and a non-trivial repulsive fixed point $g^*=-\frac{1}{3}$ with $\beta(g^*)=-1$. The exact critical point is $g^*=-\frac{1}{4}$ and the exact scaling variable for $m=3$ and $q=1$ is $N v_1^{3/2}$ instead of here $v_1 N$, where $ v_1\propto g-g^* $, in semi-quantitative agreement. 
%%%%%%%%%%%%%%%%%%%%%% 
\subsection{General fixed points: Linear perturbative approximation}

We first examine the RG flow in a perturbative approximation: we assume that $V$ is close to the Gaussian term and linearize the log term, substituting
$$\ln2V'(\rho)\mapsto 2V'(\rho)-1\,.$$
In this approximation, the flow equation reduces to the linear equation
$$\lambda{\d \over \d\lambda}V(\rho,\lambda)=V(\rho,\lambda)+\left[1- \bigl(1+  \gamma (\lambda)\bigr)\rho \right]V'(\rho,\lambda) -\ud\,, $$
or, after differentiating with respect to $\rho$ ($R=1/2V'$),   
$$\lambda{\d \over \d\lambda}R(\rho,\lambda)=\gamma(\lambda)R(\rho,\lambda)+
\left[1- \bigl(1+  \gamma (\lambda)\bigr)\rho \right]R'(\rho,\lambda).$$
The fixed point equation is simply
$$V(\rho)+[1-(1+\gamma)\rho]V'(\rho)-\ud=0 $$
and the solution is
$$V(\rho)=\ud-\ud[1-(1+\gamma)\rho]^{1/(1+\gamma)}.$$
Since $V(\rho)$ must be a regular function, we conclude
$$\gamma={1\over m}-1$$ 
with $m$ a strictly positive integer and thus
$$V(\rho)=\ud-\ud(1-\rho/m)^m.$$
The stability properties of the fixed point are determined by the eigenvalues of the operator obtained by varying the fixed point equation with respect to $V$.\par
Defining then
$$\Omega=1+(1-\rho/m){\d \over\d\rho},$$
we can write the equation for an eigenvalue $\kappa$ and eigenvector $h$ as
$$\Omega h=\kappa h+\ud\delta\gamma\rho[1-\rho/m]^{m-1},$$
where $\delta \gamma$ is determined by the condition $h'(0)=0$. Since the global normalization of $h$ is arbitrary, we can choose $\delta\gamma=2/m$.
A special solution of the equation then is
$$h_0={1\over\kappa}(1-\rho/m)^m+{m\over 1-m\kappa}(1-\rho/m)^{m-1}.$$
The solution $h_1$ of the homogeneous equation  is
$$h_1={1-\kappa\over \kappa(1-m\kappa)}(1-\rho/m)^{m(1-\kappa)}.$$
The solution $h=h_0+h_1$ must be regular at $\rho=m$ and this implies
$$\kappa=1-p/m$$
with $p$ a strictly positive integer. However, no regular solution exists for $\kappa=0$ and $\kappa=1/m$. The spectrum can be compared to the exact spectrum.\par
The value $m=1$ corresponds to the Gaussian fixed point. All eigenvalues are negative and the fixed point is stable. Moreover, the approximation corresponds to linearizing the flow equation near the fixed point and, thus, the result is also valid for the full equation.\par
The value $m=2$ corresponds to the first critical point.
Then, $\kappa=1$ is the only positive eigenvalue, corresponding to a direction of instability and all others $\kappa=-\ud,-1,-\frac{3}{2}$... are negative. More generally, for generic $m$, there exists $(m-1)$ positive eigenvalues corresponding to  $(m-1)$ directions of instability.
%%%%%%%%%%%%%%%%%%%
\subsection{Saddle point evolution: A source of difficulties}

We now return to the general flow equation \eqns{\eIRGUflowgen}. We assume that for the saddle point corresponds to the finite value $\rho=\rho_c(\lambda)$ and for $\rho-\rho_c(\lambda)\to0 $,\sslbl\ssIntRGsaddle
$$R(\rho,\lambda)=\rho+r_p(\lambda)\bigl(\rho-\rho_c(\lambda)\bigr)^p+O\left[\bigl(\rho-\rho_c(\lambda)\bigr)^{p+1}\right].\eqnd\eIntcritical $$
We  set $R=\rho+S$, which leads to the equation
$$\lambda{\d \over \d\lambda}S(\rho,\lambda)=\bigl(1+\gamma(\lambda)+S'(\rho,\lambda)\bigr)S(\rho,\lambda)- \gamma(\lambda) \rho S'(\rho,\lambda)  .\eqnd\eIRGSflowgen $$
We then set ($p\ge 1$)
$$S(\rho,\lambda)=\sum_{q=p}^\infty r_q (\rho-\rho_c)^q.$$
Identifying the term of order $(\rho-\rho_c)^{p-1}$ in the flow equation, one obtains
$$\lambda{\d \over \d\lambda}\rho_c=\gamma \rho_c\,.$$
Thus,
$$\ln\rho_c(\lambda)=\ln\rho_c(1)+\int_1^\lambda\d \lambda'{\gamma(\lambda')\over \lambda'}.$$
First, we conclude that, as expected, the critical behaviour  is invariant under the RG flow. 
However, the location $\rho_c$ {\it can have a finite non-vanishing limit}  only if $\gamma(\lambda)\to 0$ for $\lambda\to\infty$ and, thus as we will show, if the flow converges toward the {\it Gaussian fixed point}. By contrast if $\gamma(\lambda)$ has a negative limit, then $\rho_c(\lambda)\to 0$, which is a boundary value; if it has a positive limit, $\rho_c(\lambda)\to +\infty$. \par
We show later that the limit of $\gamma(\lambda)$  must be negative.
%%%%%%%%%%%%%%%%%%%%%%

\subsection {Full flow equation: Fixed points}

In terms of the function $R$, the full fixed point equation reads 
$$ \gamma R+RR'-(1+\gamma)\rho  R'=0\,.\eqnd\eIRGfixedpt  $$ 
Then, there are two cases: \par
(i) $\gamma=0$ and the equation reduces to
$$ R'(R-\rho)=0\,.$$
It has two solutions $R(\rho) \equiv {\rm const.}=R(0)$, which corresponds to the Gaussian fixed point  and  $R(\rho)= \rho$, which is  a trivial fixed point since in radial coordinates the integrand is then $N$-independent. The Gaussian fixed point has already been discussed in the framework of the linearized equation.  
\par
(ii) $\gamma\ne 0$. We then multiply the equation by $R^{-1/\gamma-2}$ and it becomes a total derivative and can be integrated. For $R (0)\ne0$ (the other case is trivial), normalizing  $R$ and $\rho$ such that $R(0)=1$, we obtain the integrated equation 
$$ R^{1+1/\gamma}(\rho)=  R (\rho)-\rho\,.\eqnd\eIRGfixpt $$ 
The function $R$ has square root singularities. The location is obtained by imposing that $R$ also satisfies the equation obtained by differentiating    with respect to $R$:
$$(1+1/\gamma) R^{1/\gamma } =1 $$
and, thus,
$$R=\left(\gamma\over1+ \gamma\right)^\gamma\,,\quad \rho={\gamma^\gamma\over (1+\gamma)^{1+\gamma}}\,.$$
The condition that solutions $V'(\rho)$ should be regular  on the real positive axis implies $\gamma<0$ and thus, according to the discussion of section \label{\ssIntRGsaddle}, $\rho_c(\lambda)\to 0$. 
%%%%%%%%%%%%%%%%%%%%%
\subsection{Duality relation and series expansion}

One can rewrite the integrated fixed point  equation as
$$R^{1/\gamma}+\rho R^{-1}=1\,.$$
Denoting   by $R(\rho,\gamma)$ the solution of the  equation, one verifies the relation 
$$ R (\rho, \gamma)=\rho R^{-\gamma}(\rho^{1/\gamma},1/\gamma ).$$
In particular, if $R( \rho,\gamma)$ is positive and regular on the real positive axis, the same applies to $R(\rho,1/\gamma)$. Moreover, it is sufficient to study the domain
$\gamma\le -1$ or $-1\le \gamma<0$.\par
%%%%%%%%
The expansion of the solution in powers of $\rho$ is
$$R(\rho)= -\gamma\sum_{n=0}{\rho^n\over n!}{\Gamma\bigl(n+\gamma(n-1)\bigr)\over\Gamma\bigl(1+\gamma(n-1)\bigr)}\ \Rightarrow R^{\alpha}=-\alpha\gamma\sum_{n=0}{\rho^n\over n!}{\Gamma\bigl(n+\gamma(n-\alpha)\bigr)\over\Gamma\bigl(1+\gamma(n-\alpha)\bigr)}\,.$$
The behaviour of the series is consistent with the location and nature of the singularities already found.

%%%%%%%
\subsection{Discussion}

One verifies easily that the fixed point solution is such that the saddle point equation $R(\rho)=\rho $ has no solution for $\rho$ finite, a result that is consistent with the analysis of the flow of the saddle point. 
Indeed, since the function $R$ has been assumed to be positive,   $R(\rho)> \rho$, and the only possible saddle points are $\rho$ infinite  or $\rho=0$. Since $\rho$ can be eliminated, the only remaining possibility is $\rho$ infinite. \par  
Asymptotically,  $R(\rho)\sim \rho$, the leading term of $V$ thus cancelling the contribution of the $\rho$-measure and the $\rho$-integral  does not converge.\par
Moreover, the leading correction is
$$R(\rho)-\rho\mathop{\sim}_{\rho\to\infty} \rho^{1/\gamma+1}.$$
The order of the fixed point is governed by $1/\gamma+1$. We seem to find a {\it continuous set of critical points}, while we know that the set is discrete. However, the interpretation of $\rho=\infty$ as a saddle point implies that the solution should be a regular function of $1/\rho$ at $\rho=\infty$. Only $\gamma=-1/n$ satisfies the condition and the fixed point equation  becomes polynomial:
$$R^n(\rho)-\rho R^{n-1}(\rho)-1=0\,.$$
One can then linearize the flow equation at the fixed point and determine the corresponding eigenvalues.
%%%%%%%%%%%%%
\subsection{A few special solutions}

With the ansatz $\gamma=-1/n$,  
the location of singularities is given by
$$\rho^n=-{n^n\over(n-1)^{n-1}}.$$
The solutions are monotonously increasing for $\rho>0$.
For complex solutions, if $R(\rho)$ is a solution, $\omega R( \rho/\omega)$ with $\omega^n=1$ is a solution.
For the first values of $n$, the equation can be solved explicitly. For example, for $n=1,2,3$,   respectively,
$$\eqalign{R&=   1+\rho \,,\quad R=\ud\left(\rho+\sqrt{\rho^2+4}\right),\cr
 R&=\left(\frac{1}{27}\rho^3+\ud+\ud\sqrt{1+\frac{4}{27}\rho^3 }\right)^{1/3}+{\rho^2 \over 9 \left(\frac{1}{27}\rho^3+\ud+\ud\sqrt{1+\frac{4}{27}\rho^3 }\right)^{1/3}}+ {\rho\over 3}\,  .\cr}$$

\subsection Non-linear renormalization

With a simple rescaling of the integration variable $\rho$, it is impossible to generate a non-Gaussian fixed point corresponding to a saddle point at a finite value of $\rho$. Non-linear changes of variables are required:
$$\rho\mapsto \rho-{1\over N}h(\rho), \ h(0)=0\,.$$
The variation (up to a possible constant) of the function $V$ is then 
$$N\delta V(\rho)= V(\rho)- h(\rho) V'(\rho)+\ud \ln 2V'(\rho)+O(N^{-2}).\eqnd\eIRGnonlin $$
The fixed point equation for $R(\rho)=1/2 V'(\rho)$  then reads
$$ \bigl(1-h'(\rho)\bigr)R(\rho)-R'(\rho)\bigl(R(\rho) -h(\rho)\bigr)=0\,.$$
If we impose the saddle point position $\rho=1$, we have the constraint
$$1-h'(1)-R'(1)\bigl(1-h(1)\bigr)=0\,.$$
We can then invert the process and, given a fixed point candidate, determine from the equation a function $h(\rho)$. 
It is obtained by solving
$$-R h'+ R' h+R -RR'=0\,.$$ 
which integrated yields
$$h(\rho)=R(\rho)\left[  \int_{0}^\rho {\d\rho'\over R(\rho')} -   \ln R(\rho)+\ {\rm const.}\right],$$ 
where the integration constant is determined by the condition $h(0)=0$.\par
Therefore, provided one chooses as a fixed point function $R(\rho)$, a function that does not vanish for $\rho\ge 0$, one can  find a a suitable change of variables. 
However, it is not clear how to guess a suitable non-linear change of variables {\it a priori}\/ in the more general situation  we examine later.
%%%%%%%%%%%%%%%%%%
\section{Quantum mechanics}

We now generalize the preceding analysis to path integrals. We consider  the free energy given by the path integral\sslbl\ssQMlargeN 
$$\e^{Z_N}=\int[\d {\bf x}(t)]\,\e^{-{\cal S}({\bf x})},\eqnd\epartmq $$
where ${\cal S}$ is an $O(N)$ invariant Euclidean action of the form
$$ {\cal S}({\bf x})=N\int\d t\left[\ud  \dot{\bf x}^2(t)+V\bigl({\bf x}^2(t) \bigr) \right]  ,\eqnd\eactmq $$
 for example, 
$$ V({\bf x})=\ud {\bf x}^2 +{g\over 4}\bigl({\bf x}^2 \bigr)^2.$$ 
We could discuss this problem using radial coordinates from the Schrödinger
equation point of view. Because eventually we want to extend the analysis to field
theory we use a more general method.\par
The large $N$ expansion can be obtained from standard
techniques   (for a review see, for example, Ref.~\refs{\rMMZJ}).
To determine the large $N$ limit, one introduces two paths $\lambda(t)$ and $\rho(t)$ and adds to the action
$$\ud N\int\d t\,\lambda(t)\left[{\bf x}^2(t)-\rho(t)\right].$$ 
The integration over the Lagrange multiplier $ \lambda(t)$ imposes the constraint
$\rho(t)={\bf x}^2(t)$  and one can  replace $V({\bf x}^2)$ by $V(\rho)$. The action then takes the form
$${\cal S}({\bf x},\lambda,\rho)=N\int \d t \left[\ud  \dot{\bf x}^2(t)  +
V\bigl(\rho(t)\bigr) +\ud \lambda(t)\left({\bf x}^2(t)-\rho(t)\right) \right].$$ 
\par
After the Gaussian integration over $\bf x$, the path integral becomes
$$\e^{{\cal Z}_N}\propto \int [\d \lambda(t)\d\rho(t)]\,\e^{-{\cal S} (\lambda,\rho)}  $$
with 
$${\cal S} (\lambda,\rho)={N \over 2}\left\{ \int \d t\left[
-\lambda (t)\rho(t)+2V\bigl(\rho(t)\bigr)\right]+\tr \ln \left(-\d_t^2+  \lambda\right)\right\}.\eqnd\eIntZNefact $$ 
The dependence on $N$ of the partition function is now explicit. 
In the large $N$ limit  the path integral can  be calculated by the steepest
descent method.  We look for two constants $\lambda,\rho$ solutions of
$$2V'(\rho)-  \lambda=0\, ,\quad \rho={1 \over 2\pi}\int {\d \omega \over
\omega^2 +\lambda}={1 \over 2\sqrt{ \lambda}}.$$
Eliminating $\lambda$ between the two equations, we obtain
$$ 8\rho^2 V'(\rho)=1\,,$$
an equation identical to the zero dimension equation, $R(\rho)=\rho$, if we now set
$$V'(\rho)={1\over 8R^2(\rho)}\,.\eqnd\eQMUdef $$  
We expect a critical point when the equation has a double root. The condition is
$$R'(\rho)=1\,.$$
One verifies that the equation expresses that the determinant of second partial derivatives  of the action for constant paths   vanishes.   Critical points are then associated with the vanishing of the determinant of the second functional derivatives at the saddle point.  The matrix of second derivatives in the Fourier representation is 
$${\bf M}={N\over2}\pmatrix{2V''(\rho)&-1 \cr -1 &-\tilde B \cr}$$
where 
$$\tilde B(\omega)={1 \over 2\pi}\int
{\d \omega'\over \left(\omega'{}^2+ \lambda\right)\left[\left(\omega-\omega'
\right)^2 + \lambda\right]} .  $$ 
In particular,
$$\tilde B(0)={1\over 4 \lambda^{3/2}}.$$
Then, using the saddle point equations, one finds
$$\det{\bf M}(\omega=0)=-\frac{1}{4}N^2\left[1+2 V''(\rho)\tilde B(0)\right]=-\frac{1}{4}N^2\left[1- R'(\rho)\right] .$$
Diagonalizing the matrix $\bf M$, we infer that for a linear combination
$\mu$ of  $\lambda$ and $\rho$, the potential vanishes in the $N\to\infty$ limit.  The path
corresponding to the second eigenvector is not critical in this sense (in one dimension a phase transition is impossible) and can be integrated out.
%%%%%%%%%%%
\subsection The $\mu$-path effective action

The resulting action is non-local but   can be expanded  in powers of $\mu$ and derivatives. 
Rescaling time and path, 
$$t\mapsto \alpha t\,,\quad \mu \mapsto {\rm const.} \times (\alpha/N)^{1/2}\mu\,,$$
 we can then look for a scaling limit \refs{\rZJONscaling}. If we relate $\alpha$ and $N$ by
$$\alpha=(Nz)^{1/5}\,,\quad z\  {\rm fixed} \ \, $$
we find that all terms in the action with more than two derivatives or powers of $\mu$ higher than three vanish with $N$. The  action then reduces to
$${\cal S}(\mu)=\ud  \int\d t\left[\dot\mu^2(t) + \sqrt{z} \mu^3(t)\right].$$ 
More generally, we can  adjust the potential $V$ in such a way that the coefficients of  all
interactions up to the $\mu^{m-1}$ vanish.  The scaling limit
corresponds then to choose
 $$\alpha=(Nz)^{(m-2)/(m+2)}\ {\rm and\ thus } \ \mu\mapsto\mu N^{-2/(m+2)}  .$$ 
Correspondingly, the effective action in the large $N$ limit becomes
$${\cal S}(\mu)=\ud  \int\d t\left[\dot\mu^2(t) + z^{(m-2)/2} \mu^m(t)\right].$$ 
Relevant perturbations correspond to adding to the action in the initial normalization a term 
$$  N v_q\int\d t\,\mu^q(t).$$
  A non-trivial scaling limit is then obtained by choosing
$$v_q\propto N^{-2(m-q)/(m+2)}.$$
By contrast, note that in matrix models the situation is more complicate  because
logarithmic deviations from a simple scaling law are found. 
%%%%%%%%%%%%%%%%%%%%%
\subsection RG inspired method

Like in the case of the simple integral, we now integrate over only one component of the path. We start from the action with $(N+1)$ components $({\bf x},y)$,
$${\cal S}_{N+1}({\bf x},y)=(N+1)\int\d t\left[\ud  \dot{\bf x} ^2+V({\bf x}^2)+\ud \dot y^2+V'({\bf x}^2)y^2+O(y^4)\right],$$
where we have assumed
$$V(\rho)=\ud\rho+O(\rho^2).$$
In the large $N$ limit, the integral over $y$ reduces to a Gaussian approximation  and (with a suitable normalization of the path integral)  one obtains
$${\cal S}_{N+1}({\bf x})=(1+1/N){\cal S}_N({\bf x})+\ud W +O(1/N)   $$
with
$$W=\tr\ln\left(-\d_t^2+2V'({\bf x}^2)\right).$$
Immediately we face a new problem, the contribution $W$ to the action is not local. 
Then we rescale ${\bf x}$,
$${\bf x}\mapsto {\bf x}(1-(1+\gamma)/2N),$$
even though we have seen in the zero-dimensional example that a linear rescaling may lead to difficulties. We then obtain the variation of the action
$$N\delta {\cal S}_N({\bf x})= \int\d t\left[-\ud \gamma \dot{\bf x} ^2+V({\bf x}^2) -(1+\gamma) {\bf x}^2V'({\bf x^2}) +\ud W\right] . $$
We still have to perform a time renormalization to normalize the coefficient of $\dot {\bf x}^2$ to $\ud$:
$$t\mapsto t(1-\gamma/N).$$
This yields an additional contribution $-\gamma V$:
$$N\delta {\cal S}_N({\bf x} )= \int\d t\left[ (1-\gamma)V({\bf x}^2) -(1+\gamma) {\bf x}^2V'({\bf x^2}) +\ud W\right] . $$
We can then determine $\gamma$ by the condition $\delta V'(0)=0$.\par
A first way to solve the locality issue is to expand $W$ at leading order in $V'-1$,
$$W=W_0+\ud\int\d t\left[V'({\bf x}^2)-\ud\right]+O\bigl((V'-\ud)^2\bigr),$$
where $W_0$ is a constant but the next term in such an expansion breaks locality. 
\smallskip
{\it The local approximation.}
More globally, from the preceding analysis, we expect that, for $N$ large, a local expansion (which is also a large $V$ expansion) should be meaningful because the fluctuations of ${\bf x}^2$ are small. Then,
$$\tr\ln\left(-\d_t^2+2V'({\bf x}^2)\right)\sim\int \d t\sqrt{2V'({\bf x}^2)},\eqnd\edetlocal $$
the leading correction being 
$$\int\d t\left({\bf x}\cdot\dot{\bf x}\right)^2{V''{}^2({\bf x}^2)\over
2\left[2V'({\bf x}^2)\right]^{5/2}}.$$
The flow equation then becomes a simple scalar equation for the function $V(\rho)$ of the form
$$N{\partial \over \partial N}V(\rho)=(1-\gamma)V(\rho)-(1+\gamma)\rho V'(\rho)+\ud \left[\sqrt{2V'(\rho)}-1\right],\eqnn $$
where the constant is adjusted to enforce consistency for $\rho=0$.\par
Differentiating with respect to $\rho$ and setting $V'(\rho)=1/8 R^2$, one obtains
$$N{\partial \over \partial N}R(\rho) =  \gamma R(\rho)- (1+\gamma)\rho R'(\rho)+ R(\rho)R'(\rho),\eqnn $$
an equation identical to equation \eqns{\eIRGUflowgen} after the change $N\mapsto N\lambda$.
%%%%%%%%%%%%%%%%%
\section{Fixed point equation}

Like in the  example of the simple integral, we first consider the perturbative and also linear approximation. Alternatively, we then use the first term of the local expansion of $W$. 
%%%%%%%%%%%%%%%%%%%%%%%%%%%
\subsection{The perturbative approximation: Fixed points}

The fixed point equation then reads ($\rho={\bf x}^2$)
$$   2(1-\gamma)V(\rho)+ [1-2(1+\gamma) \rho ]V'(\rho)-\ud =0\, . $$
The solution is
$$V={1\over 4(1-\gamma)}\left[1-\bigl(1-2(1+\gamma)\rho\bigr)^{(1-\gamma)/(1+\gamma)}\right].$$
The regularity condition implies
$$\gamma={1-m\over 1+m}\ \Rightarrow\ V={1+m\over8m}\left[1-\left(1-{4\rho\over m+1}\right)^m\right]\,, $$
where $m$ is a strictly positive integer.
%%%%%%%%%%%%%%%%%%%%%%%%%%%%%
\subsection{The perturbative approximation: Eigenvalues}

We define the operator
$$\Omega={4m\over m+1}+\left(1-{4\over m+1}\rho\right){\d\over \d\rho}\,.$$
Denoting by $\kappa$ the eigenvalue and $h$ the eigenvector, we find the eigenvalue equation
$$\Omega h=\kappa h+2 \delta\gamma{1+m\over8m}\left[1+(m-1)\left(1-{4\rho\over m+1}\right)^m
+m\left(1-{4\rho\over m+1}\right)^{m-1} \right],$$ 
where $\delta\gamma$ is determined by the condition $h'(0)=0$.
\par
The solution $h_1$ of the homogeneous equation is
$$h_1=\left(1-{4\rho\over m+1}\right)^{m-\kappa(m+1)/4}. $$
The regularity condition implies
$$\kappa=4\,{m-q\over m+1}\,. $$
%%%%%%%%%%%%%%%%%%
\subsection{The local expansion}

In the local approximation, we reduce  the problem essentially to the example of the simple integral. The fixed point equation now reads
$$   (1-\gamma)V({\bf x}^2) -(1+\gamma) {\bf x}^2V'({\bf x^2}) +\ud\sqrt{2V'({\bf x}^2)}=0\, . $$
Setting $\rho={\bf x}^2$ and differentiating the fixed point equation with respect to $\rho$, we obtain
$$-2\gamma V'(\rho)-(1+\gamma)\rho V''(\rho)+{V''(\rho)\over 2\sqrt{2 V'(\rho)}}=0\,.$$
We then set  $V'=8/R^2 $ and the equation becomes indeed identical to the equation for the simple integral:
$$  \gamma R(\rho)+ R'(\rho)R(\rho)-(1+\gamma)\rho R'(\rho)  =0\,.$$
We do not have to repeat the discussion of the simple integral. The only modification is the implication of the fixed point solution for $V$.
\par
For example, since for $\rho\to\infty$, $R(\rho)\sim\rho$ then $V(\rho)$ goes to a constant with as a leading correction $-1/8\rho$.  

%%%%%%%%%%%%%%
\section {Statistical field theory}
 
The analysis leading to the double scaling limit can be generalized to  statistical field theories in two dimensions with generic potentials and in three dimensions to theories with a sextic potential \refs{\rZJONscaling}.\par
From 
the discussion of the quantum mechanical case, we have already some intuition
about the main difference with $d=1$. The critical limit
corresponds to having one bound state associated with the composite ${\Bfg \phi}^2
$ field becoming massless, (the ${\Bfg \phi}$-field itself remaining non-critical)
that is, to a phase transition of an Ising-like 
system. For $d=1$ a phase transition is impossible while it is possible in
higher dimensions. We  now show how this difference shows up technically,
first in the special case of the $({\Bfg \phi}^2)^2$ interaction.
We consider the partition function  
$${\cal Z}_N = \int \left[ \d {\Bfg \phi}(x) \right] \e^{-{\cal S} ( {\Bfg \phi})},$$
where ${\cal S}({\Bfg \phi})$ is a $ O(N) $ symmetric action:
$${\cal S}( {\Bfg \phi})= N\int \left\{ \ud \left[\partial_{\mu} {\Bfg \phi} (x)
\right]^{2}+V\bigl({\Bfg\phi}^2(x)\bigr) \right\} \d^{d}x \eqnd\eactON$$
with as simplest example
$$V(\rho)= \ud r \rho +\frac{1}{4}g \rho^2  .$$
A cut-off of order $1$, consistent with the symmetry, is implied. More
precisely  we have written explicitly in action \eqns{\eactON} only the two first
terms of the inverse propagator in a local expansion (in  Fourier space small
momentum) expansion. In particular, the parameter $r$ is defined as the
value of the inverse propagator at zero momentum. \par
Following the method explained in section \label{\ssQMlargeN}, 
we introduce two fields $\lambda(x),\rho(x)$,  add to the action \eqns{\eactON}, 
$$\ud N\int\d^d x\,\lambda(x)\left[{\Bfg\phi}^2(x)-\rho(x)\right] $$
and, accordingly, replace $ V({\Bfg \phi}^2)$ by $V(\rho)$.
We then perform the Gaussian
integration over ${\Bfg \phi} $ and find 
$${\cal Z}= \int \left[ \d  \lambda(x)\d\rho(x)\right]  \e^{-{\cal S} (\lambda,\rho)} 
$$
with 
$${\cal S} (\lambda,\rho) ={N \over 2} \left\{ \int   \d^{d}x\left[-\lambda(x)\rho(x)+2V\bigl(\rho(x)\bigr)\right] +\tr\ln \bigl( -\Delta_x+   \lambda   \bigr) \right\} . $$
In the large $N$ limit the field integral can be calculated by the steepest
descent method. 
%%%%%%%%%%%%%
\subsection{The large $ N $ limit}

We look for a uniform saddle point $\lambda(x)=\lambda$, $\rho(x)=\rho$.
The saddle point equations are
$$ -\lambda + 2V'(\rho)=0\,,\quad -\rho+{1 \over (2\pi)^{d}} \int^\Lambda{ \d ^{d}p \over p^{2} +  \lambda
}  = 0\, .\eqnd\eNQFTsaddle  $$ 
where we have introduced a cut-off $\Lambda$ since otherwise the integral is divergent for $d\ge 2$. With a cut-off and for $d<4$, the second equation becomes
$$-\rho+{1\over(4\pi)^{d/2}}\Gamma(1-d/2)\lambda^{d/2-1}+\Lambda^{d-2}C(d)=0\,,$$
where the pole at $d=2$ of $C(d)$ cancels the pole of the $\Gamma$-function. We call $\rho_0(d)$ the cut-off dependent term, set 
$$K(d)={\Gamma(1-d/2)\over(4\pi)^{d/2}},\eqnd\eNQFTKdef $$
and introduce the function
$$R(\rho)=  K(d) [2V'(\rho)]^{ d/2-1} \,.\eqnd\eQFTUdef $$
Using the first equation \eqns{\eNQFTsaddle}, the equation can then be rewritten as
$$\rho-\rho_0=R(\rho).\eqnd\eNQFTsaddlea $$ 
The critical point is defined by 
$$0=1-R'(\rho)=1+2V''(\rho) {1\over(2\pi)^d}\int{\d^d q\over (q^2+\lambda)^2 } \,.\eqnd\eQFTcritft$$
The determinant of the matrix of the second derivatives of the action at the saddle point,
$${\bf M}= { N\over2} \pmatrix{-\tilde\Delta(p)& -1 \cr -1 & 2V''(\rho)\cr} $$
with
$$\tilde\Delta(p)={1\over(2\pi)^d}\int{\d^d q\over (q^2+\lambda)\bigl((p+q)^2+ \lambda\bigr)},$$
 is proportional to
$$1+2 V''(\rho)\tilde\Delta(p).$$
The criticality condition \eqns{\eQFTcritft} thus implies that the determinant vanishes at $p=0$ and the propagator of a linear combination $\mu$ of the $\rho$ and $\lambda$ fields,
$$\mu(x)=\rho(x)-\rho+2V''(\rho)\bigl(\lambda(x)-\lambda\bigr),$$
 has a pole at zero momentum. The  field $\mu(x)$ becomes massless while the other component and also the ${\Bfg \phi}$-field remain  massive.
%%%%%%%%%%%%%%%%%%%%%%%%%%%%%%
\subsection{The scaling limit in field theory}

We have determined the partition function in the large $N$ limit. We now look for a scaling limit \refs{\rZJONscaling}.
Since the $\mu$ field is a one-component field, it can remain critical
for $d>1$ even in presence of interactions. We have to examine the most IR
divergent terms in perturbation theory. We face a standard problem in the
theory of critical phenomena.
The deviation
$\varepsilon $ of a parameter in $V$ from its critical value
plays exactly the role of a deviation from the critical temperature. 
\medskip
{\it Two dimensions.} We first examine dimension 2. The effective action for the $\mu$-field is
non-local and contains arbitrary powers of the field. However, because the
other fields are not 
critical,  we can again make a local expansion. Standard arguments of the
theory of critical phenomena tell us that the most IR divergent terms come
from interactions without derivatives and with the lowest power of the field.
Here the leading interaction is proportional to $\mu^3$. 
To characterize the IR divergences of the perturbative expansion in powers in
$1/N$ we rescale distances and field $\mu$,
$$\mu(x) \mapsto  \mu(x) N^{-1/2},\quad x\mapsto
\Lambda x,$$
where $\Lambda$  plays the role of a cut-off. The effective action at
leading order, after some additional finite renormalizations, is
$${\cal S}_{\rm eff.}(\mu)=\int\d^2 x\left[\ud
\left(\partial_{\mu}\mu\right)^2  +\ud \varepsilon \Lambda^2\mu^2 +
{\Lambda^2 \over 6\sqrt{N}}\mu^3 \right],$$
where a cut-off $\Lambda$ is again implied. \par
If $d<4$ the theory is super-renormalizable. We want the coefficient of
$\mu^3$ to have a limit and we thus set  
$$\Lambda^2/\sqrt{N}= u.$$
In contrast with quantum mechanics, we cannot just fix the product $\varepsilon
\Lambda^2=u\varepsilon \sqrt{N} $ because the perturbative expansion is not
finite in the large $N$, large cut-off $\Lambda$ limit. Instead, we have to
introduce a counter-term which renders $\langle\mu(x)\rangle$ finite. A short
calculation yields the relation 
$$\varepsilon={1 \over u\sqrt{N}}\left[{1 \over 2}\mu^4+{u^2 \over 16\pi}
\ln\left( Nu^2/m^4\right) \right],$$
in which $m$ is a renormalized mass parameter. We have shown that a scaling
limit exists which leads to a renormalized $\mu^3$ field theory. However, the
relation between $ \varepsilon$ and $N$  has itself no longer a simple
scaling form.
\medskip
{\it Higher dimensions.} The situation in higher dimensions is similar,
though slightly more complicate,   as
long as the theory is super-renormalizable, because more counter-terms are required. Naive scaling would predict a
scaling variable $N \varepsilon^{(6-d)/2}$. However, due to counter-terms,
$\varepsilon$ does not go to zero as fast as naively expected. We 
perform here the general analysis not for the $\mu^3$ theory but rather for the
more interesting $\mu^4$ theory, which can only been obtained from ${\Bfg \phi}$
interactions depending on more parameters.
\medskip
{\it  Multicritical points.} The problem can be dealt with
by the method explained in the case of quantum mechanics. The results are
very similar. Because only one mode is critical, one can 
introduce additional parameters in the effective interaction of the critical
mode in such a way that the most IR relevant terms can be cancelled. In the
language of critical phenomena we reach multicritical points. We generate then
renormalized $\lambda^p$ interactions provided we again choose the relation
between $\varepsilon$ and $N$ to cancel the UV divergences of perturbation
theory. In two dimensions the relation, with standard normalizations, is 
$$\Lambda^2=Nu,\quad \varepsilon={1 \over Nu}\left(m^2+{1\over8\pi}\ln
(Nu/m^2) \right).$$
In dimension $2<d<4$ the relation becomes more complicate because more
counter-terms are needed. Note that at leading order for $N$ large
$$\varepsilon\sim {\Lambda^{d-2} \over N}\sim N^{-(4-d)/2},$$
while naive scaling would indicate 
$$\varepsilon\sim N^{-2/(4-d)}.$$
In dimensions $d>4$ perturbation theory is no longer IR divergent and,
therefore, no scaling limit can be found. 
%%%%%%%%%%%%%%%%%%%%
\subsection RG inspired strategy

We now generalize the strategy used in the quantum mechanics case ($d=1$) to field theory. Again we evaluate the determinant generated by the integration over one component in the local approximation. We start from the action \eqns{\eactON} for $(N+1)$ components and integrate over only one component in the large $N$ limit:
$$\eqalignno{{\cal S}_{N+1}({\Bfg \phi},\chi)&=(N+1)\int\d^{d}x  \left\{ \ud \left[\partial_{\mu} {\Bfg \phi} (x)
\right]^{2}+V\bigl({\Bfg\phi}^2(x)\bigr)\right\}\cr
&\quad+\int\d^{d}x  \left\{  \ud\bigl(\partial_\mu\chi(x)\bigr)^2+\chi^2(x)V'\bigl({\Bfg\phi}^2(x)\bigr)+\cdots \right\}& \eqnd\eactFTRGON \cr}$$
Performing the integration over $\chi$ in the Gaussian approximation, one finds the action
$$(1+1/N){\cal S}_N({\Bfg\phi})+\ud\tr\ln[-\Delta_x+2 V'({\Bfg\phi}^2)] +O(1/N).$$
In the local approximation, the determinant can be evaluated and one finds 
$$  \tr\ln[-\Delta_x+2 V'({\Bfg\phi}^2)] \sim  {2K(d)\over d}\int \d ^{d}x   \left[2V'\bigl({\Bfg\phi}^2(x)\bigr) 
\right]^{d/2}, \eqnd\eNdetSFT  $$
where $K(d)$ has been defined in equation \eqns{\eNQFTKdef}. 
In even dimensions, an additional cut-off dependent contribution cancels the pole of the $\Gamma$-function. For $d<4$, it is proportional to $V'({\Bfg\phi}^2)$ and for $d=4$ it is quadratic in $V'({\Bfg\phi}^2)$. Adding to expression \eqns{\eNdetSFT} the divergent part, one obtains
$$\int \d ^{d}x \left\{ { K(d)\over d} \left[2V'\bigl({\Bfg\phi}^2(x)\bigr) 
\right]^{d/2} + \rho_0(d)V'\bigl({\Bfg\phi}^2(x)\bigr)\right\} .$$
We now introduce a field renormalization
$${\Bfg \phi}(x)\mapsto{\Bfg \phi}(x)\bigl(1-(1+\gamma)/N\bigr).$$
Omitting the cut-off dependent term, the variation $\delta{\cal S}_N$ of the action   then becomes
$$\eqalign{&N\delta{\cal S}_N({\Bfg\phi})\cr&\  =\int\d^d x\left\{-\ud\gamma(\partial_\mu{\Bfg\phi})^2+ V({\Bfg\phi}^2)-(1+\gamma){\Bfg\phi}^2V'({\Bfg\phi}^2)+ {K(d)\over d}\left[2V' ({\Bfg\phi}^2 )  
\right]^{d/2}\right\}. \cr} $$
A space rescaling is then required to normalize the coefficient of the kinetic term to $\ud$:
$$x\mapsto x\left(1+{\gamma \over (d-2)N}\right),$$
{\it a form that shows that $d=2$ has to be examined separately}. The final form of the variation of the potential then reads
$$N\delta V(\rho)=(1+\gamma d/(d-2))V(\rho)-(1+\gamma)\rho V'(\rho)+ {K(d)\over d}[2V'(\rho)]^{d/2} +\rho_0 V'(\rho). $$
Differentiating with respect to $\rho$, one obtains
$$N\delta V'(\rho)={2\gamma \over d-2}V'(\rho)-(1+\gamma)\rho V''(\rho)+ V''(\rho) \left[K(d)(2V'(\rho))^{d/2-1}+\rho_0\right]. $$
In terms of the function \eqns{\eQFTUdef},
$${\delta R\over R}=(d-2)  {\delta V'\over2 V'} $$
and
$${V''\over V'}={2R'\over(d-2)R}\,.$$
The variation equation for $R$ then reads
$$\delta R=\gamma R -(1+\gamma)\rho R'+ (R+\rho_0) R' \,,\eqnn $$
which leads to an equation identical to equation \eqns{\eIRGUflowgen} up the divergent addition.
\smallskip
{\it Fixed point equation.}
For $\gamma\ne0$, we then multiply the fixed point equation
$$\gamma R -(1+\gamma)\rho R'+ (R+\rho_0) R' =0$$
 by $R^{-1/\gamma-2}$ and integrate:
$$\rho-{\rho_0\over 1+\gamma}-R(\rho)=R^{1/\gamma+1}(\rho)R^{-1/\gamma-1}(0).$$
We find the same equation as in previous examples, up to a simple shift of $\rho$.\par
The equation is no longer singular at $d=2$ but a more detailed study of the $d=2$ limit is required.
%%%%%%%%%%%%%%%%%%%%%%%%%%%%%%%%%%%%%%%%%%%%%%%%%%%%%%%%%%%%%555
\section{The Hermitian one-matrix model and random surfaces}

The matrix model representation of two-dimensional
quantum gravity \refs{\rDKAJ}  has led to explicit solutions for
minimal
conformal fields coupled to gravity \refs{\rBKDSGM, \rBKZGP, \rPDFPGZJ}. However, this
approach has not been of real help for understanding the
difficulties which arise when the central charge $ c $ of the matter
field is larger than one although it is very easy to write matrix
models for those cases as well. \par
We  note that the central result of matrix models 
is the existence of a `double scaling limit'  that is, a continuum 
limit with critical exponents that describe how the  coupling 
constants of the theory have to be tuned to reach this limit. 
These simple scaling laws for $ c<1 $, with logarithmic deviations at 
$ c=1$ are quite reminiscent of the
theory of continuous phase transitions for space dimensions $d\le 4$. 
  There, information about the critical behaviour is obtained without solving models explicitly but constructing a renormalization group \refs{\rKGWJK}. This is a strategy we want to use again here, extending the approach of reference \refs{\rEBZJ} and in the spirit of the method we have explained in the vector model example in section \label{\ssIntRG}.
  %%%%%%%%%%
\subsection{The one-matrix models}

We consider an integral over $N\times N$ Hermitian matrices $M$ of the form  
$$\e^{Z(N) }=\int\d M\,\e^{-N\tr V(M)},$$
where  $V$ is a polynomial, for example, 
$$V(M)=\ud M^2 +\frac{1}{3} g M^3.$$
The formal expansion of $Z(N,g)$ in powers of $g$  generates a sum of connected {\it fat} Feynman diagrams (see figure \label{\figmatrix})  whose dual are {\it connected triangulated surface}  (see figure \label{\figrandtri}). Giving a unit area to all triangles, one concludes that the term of order $g^n$ is the sum of surfaces of area $n$.  Moreover, $Z(N,g)$ has the large $N$ expansion 
$$Z (N,g)=N^2 Z_0(g)+Z_1(g)+{1\over N^2}Z_2(g)+\cdots =\sum_h N^{2-2h} Z_h(g),$$
where $Z_h(g)$ is the sum of all surfaces of genus $h$.
%%%%%%%%%%%%%%%%%%%%%%%%%%%%%%%%%%%%%%%%%%%%%%%%%%%%%%%%%%
\topinsert
\epsfxsize=49.00mm
\epsfysize=25.mm
\kern-6mm
\centerline{\kern8mm\epsfbox{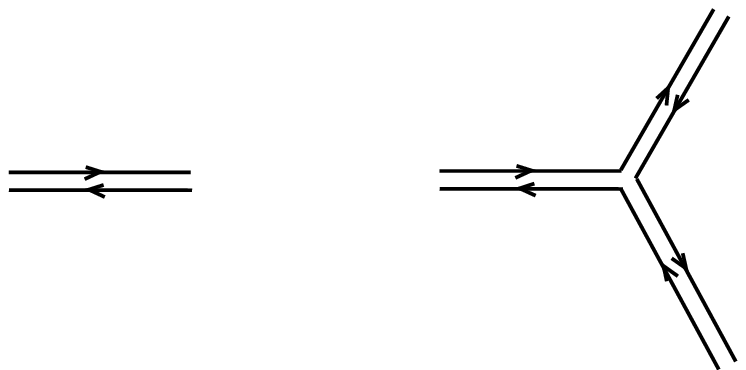}}
\figure{0.1mm}{Hermitian matrix propagator.  Hermitian matrix three-point vertex.} 
\endinsert
\figlbl{\figmatrix}
\topinsert
\epsfxsize=46.mm
\epsfysize=30mm
\centerline{\epsfbox{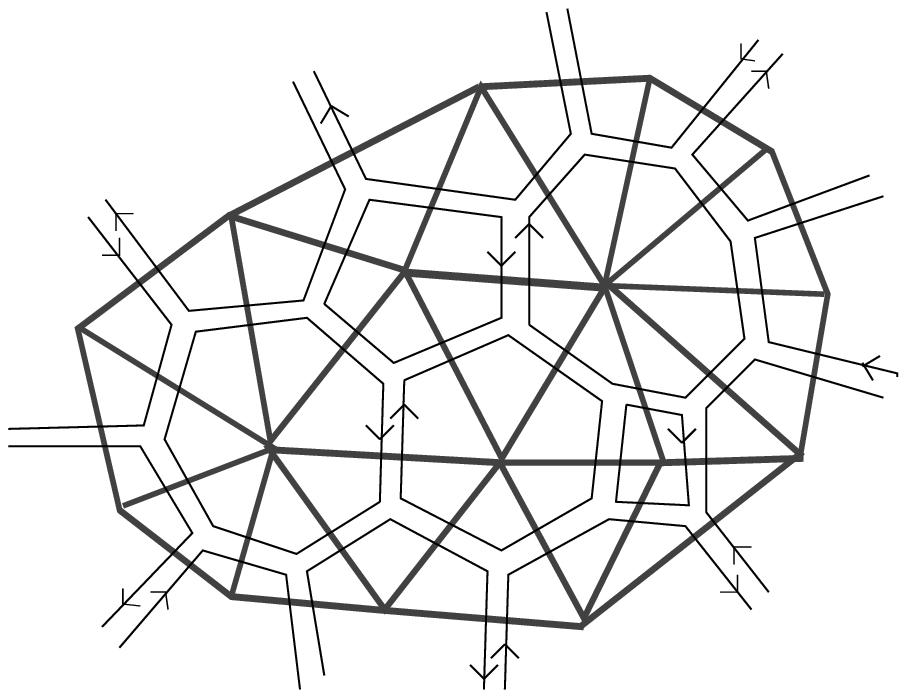}}
\figure{0.0mm}{$\tr M^3$  vertex and triangulated surface.}
\endinsert
\figlbl{\figrandtri}
%%%%%%%%%%%%%%%%%%%%%%%%%% 
\subsection{The continuum limit}

The quantity $Z_0(g)$ can be calculated by the steepest descent method. One then finds a critical value $g_c$ of $g$ where $Z_0(g)$ is singular,
$$Z_{0,{\rm sing.}}(g)\propto (g-g _c)^{2-\gamma}.$$
The specific heat exponent $\gamma$ is called {\it string susceptibility} in the string terminology. At this point the average surface area $A$ diverges:
$$A=\left<n\right> ={\partial\ln Z_{0,{\rm sing.}}(g)\over \partial g}\propto{2-\gamma\over g-g_c}\,,$$
allowing to define a {\it continuum limit} by shrinking the size of the polygons to zero. 
%%%%%%%%%%%%%%%%%%%%%%%%%%%%%%%
\subsection{The double scaling limit}

More generally, it can be shown that $Z_h(g)$ diverges at the same critical value like 
$$Z_h(g)\propto (g_c-g)^{(2-\gamma)(1-h)}.$$
This allows defining a double scaling limit in which $N\to\infty$ in a correlated way with $g\to g_c$. Setting
$$\kappa^{-1}=N(g-g_c)^{(2-\gamma)/2}$$
one finds,  in the {\it double scaling limit}, a sum over continuum surfaces with increasing genus $h$:
$$Z(N,g)\sim Z(\kappa)=\sum_{h=0} \kappa^{2-2h}f_h\,.\eqnd\escalingform $$ 
In the double scaling limit, the partition function can be calculated explicitly by various techniques, for example, using orthogonal polynomials \refs{\rPDFPGZJ}. 
%%%%%%%%%%%%%%%%%%%%%%%
\subsection{Multicritical points and central charge}

Up to now we have assumed a simple model in which $V(M)$ depends only on one parameter. If $V(M)$ depends on several parameters, like in the theory of critical phenomena one can tune some of them to reach {\it multicritical points} with {\it different} values of $\gamma$ and {\it scaling partition functions}. The continuum Liouville theory predicts a relation between $\gamma$ and the {\it central charge} $c$ that labels the unitary discrete series of conformal 2D field theories,
$c=1-6/m(m+1)$:
$$\gamma=\frac{1}{12}\left(c-1-\sqrt{(1-c)(25-c)}\right)=-\frac{1}{m}\,.$$ 
%%%%%%%%%%%%%%%%%%%%%%%%%%%
\subsection{The renormalization group approach}

Again, instead of trying to solving models exactly, we want to construct a renormalization group. If we are able to construct such a renormalization group,  we know that the scaling laws and the exponents, which 
characterize the double scaling limit, will follow automatically.\par
Therefore, we have to understand how the coupling constants of 
the theory, ---in the simplest example the string coupling constant and the cosmological 
constant  mapped, respectively, onto the size $ N $ of the matrices and 
some matrix parameter,---  evolve under a rescaling of the 
regularization length introduced in the triangulation of the 
world sheet.\par
We  expect that a change $ N \mapsto N+\delta N $ can
be compensated by a change of the matrix parameter with the same
continuum physics. We show that for matrix models this is indeed the case \refs{\rEBZJ} (see also Ref.~\refs{\rmatrg}), at least in some approximation scheme.

\subsection{Construction of flow equation: Gaussian integration}

In order to study the existence at once of critical and multicritical 
points within one-matrix models, we immediately allow for an
arbitrary analytic even potential depending on an $N\times N$ matrix ${\Bfg\phi}_N$ matrix,
$$ V ({\Bfg \phi}_ N ,g_k) =N\tr V({\Bfg \phi}_N)\equiv N \sum^{ \infty}_ 1{g_k \over 2k} \tr 
{\Bfg \phi}^{ 2k}_N  \eqnd\egenmatpot $$
with $ g_1=V''(0)=1 $ and we set  $g_2=V^{(4)}(0)/6= g$.  \par
We then consider the matrix integral 
$${\cal I}_ N(g) = \int \d  {\Bfg \phi}_ N\,\e^{-N \tr V({\Bfg \phi}_ N)}.$$
To calculate the integral over $(N+1)\times (N+1)$ matrices $ {\Bfg \phi}_{ N+1} $,
we parametrize the matrix $ {\Bfg \phi}_{ N+1} $ in terms of an $ N\times N $
submatrix $ {\Bfg \phi}_ N $, 
a complex $ N $-component vector $ v_a  $, and a real number $ \alpha$:
$${\Bfg \phi}_{ N+1} = \left( \matrix{\displaystyle {\Bfg \phi}_ N &
\displaystyle v_a 
\cr\displaystyle v^\ast_a & \displaystyle \alpha \cr} \right) .\eqnd\ephiNun  $$
One verifies easily that all the terms involving $ \alpha $ are of 
relative order $ 1/N $ and can be dropped at leading order; we 
thus can set $ \alpha =0 $. 
Expanding the polynomial $V$ in powers of $v$, we obtain
$$ \tr V({\Bfg \phi}_{N+1}) = \tr V({\Bfg \phi}_N) +   v^* V'({\Bfg \phi}_N){\Bfg \phi}_N^{-1} v+O
\left(|v|^4 \right) .$$ 
In the large $N$ limit, at leading order we truncate the expansion at {\it the  quadratic order} in $\bf v $, integrate over $v$ to find the new action
$$ V({\Bfg \phi}_{N}) + \delta V({\Bfg \phi}_{N})    = (N+1)   \tr V({\Bfg \phi}_N)  +  \tr\ln\left[V'({\Bfg \phi}_ N ){\Bfg \phi}_ N^{-1} \right]\,.   $$ 
After rescaling  of the matrix to enforce the condition $ \delta V''(0)=0$, 
 $$ {\Bfg \phi}_ N =  \zeta  {\Bfg \phi}^{ \prime}_ N \ {\rm with}\   \zeta   = 1 - { g+1/2 \over  N} + O \left({1 \over N^2} \right),   $$
the variation of the function $V$ becomes ($\mu$ is a real variable and $1/N$ is considered as a continuous variable)
$$N{\partial \over\partial N}V(\mu)\equiv N\delta  V(\mu) =  \left[V(\mu) -\left(g +\ud\right)\mu V'(\mu)+\ln\bigl(V'(\mu  )/\mu \bigr)\right].\eqnd\eonematrixflow $$
%%%%%%%%%%%%%%%%%%%%%%%%%%%%%%%%%%%%%%%%%%%%%%%%%%%%
\subsection{Perturbative approximation: the gravity exponent}

We know from the exact solution that the one-matrix $ {\Bfg \phi}^ 4 $ model is sufficient to describe, near its 
critical point,  pure gravity $ (c=0) $. It consists of the 
integral over an $ N\times N $ Hermitian matrix $ {\Bfg \phi}_ N $,
$$ {\cal I}_ N(g) = \int  \d   {\Bfg \phi}_ N\, \exp\left[- N \tr V \left({\Bfg \phi}_N,g
\right)\right]  $$ 
with an `action'
$$V ({\Bfg \phi}_ N,g ) = \tr \left(\ud{\Bfg \phi}^
2_N+\frac{1}{4}g {\Bfg \phi}^ 4_N \right).\eqnd\eactSN $$
The double scaling limit 
is then reached in the limit $g\to g^*$ with 
$$ g^* = -1/12   $$
and the susceptibility exponent is
$$ \gamma   = -1/2\,.   $$ 
In the RG approach, to reduce the flow equation to a one parameter flow, we expand the equation in powers of $\mu$, {\it a purely numerical approximation}. To order $\mu^4$, the equation reduces to
$$N{\partial g \over \partial N}\equiv -\beta(g)=-g(1+6g).$$
The $\beta$-function has two zeros corresponding to two fixed points: $ g^*  = 0 $ is the attractive Gaussian fixed point and $g^*= -1/6 $ is the
non-trivial repulsive fixed point since
$$ g^*  = -1/6 \  \Rightarrow\ \beta'(g^*)=-1\,.  $$
Indeed, the pure gravity exponents are obtained only when one  `tunes' the 
cosmological constant $ g $ near its critical value $ g^* $. In the exact theory the  value of $ g^* $
is $ -1/12 $, in contrast to our first approximation $ -1/6 $.
Since at this order $ \beta'(g^*) $ is equal to $-1$,
the calculation yields for the string susceptibility exponent \refs{\rEBZJ} 
$$ \gamma  = 2 + {2   \over \beta'(g^*)}= 0\,,    $$
to be compared with the exact value $ -1/2 $  .  
%%%%%%%%%%%%%%%%%%%%\
\subsection{Linear approximation}

We  give here only a brief discussion of multicritical points, in this RG approach,   based on assuming a {\it small deviation from the Gaussian model}.  \par
We use  an approximation of small deviation from the Gaussian model replacing the determinant contribution by (this is at variance with has been done in the first example) by the linear deviation:
$$ \tr\ln\left[V'({\Bfg \phi}_ N ){\Bfg \phi}_ N^{-1} \right]\sim \tr\left[V'({\Bfg \phi}_ N ){\Bfg \phi}_ N^{-1} -1\right].$$ 
%%%%%%%%%%%%%%%%%%%%%%%%%%%%%%%%%%%%%%%%%%%%%%%%%%%%%%%%%%%%%%55
The fixed point equation corresponding to equation \eqns{\eonematrixflow} reduces to
$$ V(\mu)-(g+\ud)\mu V'(\mu)+V'(\mu)/\mu-1=0\,. $$
Integrating we find
$$V^*(\mu)=1-\left[1-\mu^2(g+\ud)\right]^{1/(1+2g}.$$
Since $V^*$ must be regular function (and assuming $g+\ud>0$), we infer
$${1\over 1+2g_c}=m\ \Rightarrow\ g_c={1\over 2m}-{1\over 2}\,,$$
and thus 
$$V^*(\mu)=1-(1-\mu^2/2m)^m.$$
First, $ m=1 $ corresponds to the Gaussian fixed point. For $ m=2 $ one finds $ \mu^2/2 - \mu^4/16 $,
instead 
of the exact 
$ \mu^2/2 - \mu^4/48 $; for $ m=3 $, $
\mu^2/2-\mu^4/12 + \mu^6/216 $ instead of $ \mu^2/2 - \mu^4/12 + \mu^6/180 $.  \par
With this simple approximation, we find the right sequence of 
multicritical points with their property of being 
alternatively unbounded below for even $ m$'s and bounded for
odd $ m$'s. \par
For the $ m $-th 
multicritical point, critical exponents are related to the eigenvalues at the fixed point of the operator  obtained by varying $\delta V$ with respect to $V$. We define
$$\Omega= 1+\left({1\over\mu}- {\mu\over 2m}\right){\d\over\d\mu}   \,.$$
Denoting by $\kappa$ an eigenvalue, we have to solve the eigenvalue equation
$$\Omega h=\kappa h+\delta g \mu^2(1-\mu^2/2m)^{m-1}\,,$$
where $\delta g$ is determined by the condition that the eigenvectors $h$ satisfy $h''(0)=0$.\par
A special solution to the equation is
$$h_0(\mu)=2m \delta g\left[{1\over\kappa}(1-\mu^2/2m)^{m}+{m\over 1-m\kappa} (1-\mu^2/2m)^{m-1}\right].$$ 
The  solution $h_1$ of the homogeneous equation such that $h_1(0)+h_0(0)=0$ is
$$h_1(\mu)=-{2m \delta g\over \kappa(1-m\kappa)} (1-\mu^2/2m)^{ -m(\kappa-1)}.$$
The function $h_1$ must be regular for $\mu^2=2m$ and thus
$$\kappa=1-{p\over m}\,,$$
where $p$ is a positive integer, $p\ge 1$.  In general the values $\kappa=1/m$ and $\kappa=0$ are excluded because the regular solution $h_1+h_0$ vanishes. For $m=1$, $ \kappa< 0$ and the Gaussian fixed point is stable. For $m=2$, the only value is $\kappa=1$ and we recover the direction of instability. In general, one finds $(m-1)$ eigenvalues corresponding to relevant perturbations. By tuning $(m-2)$ parameters, one can select the eigenvector corresponding to the smallest eigenvalue $2/m$. For the $ m $-th multicritical point, instead of the exact value $ 3/2-m $, one thus finds  
$\gamma= 2+2/(-2/m)=2-m$, in good qualitative and semi-quantitative agreement with the exact answer (here we have calculated the opposite of the eigenvalues of the usually defined $\beta(g)$-functions). 

%%%%%%%%%%%%%%%%%%%%%%%%%%%%%%%%%%%%%%%%%%%%%%%
\subsection {General calculation}

We now keep the   exact fixed point equation  as obtained after Gaussian integration:  
$$ V(\mu)-(g+\ud)\mu V'(\mu)+\ln \bigl(V'(\mu)/\mu\bigr)=0\,.\eqnd\ematRGfixedpoint $$ 

\medskip
{\it Fixed point solution}. 
Setting (this assumes $V$ is even)
$$f(\rho=\mu^2/2)=V'(\mu)/\mu$$
and differentiating the fixed point equation, one finds
$$f'(\rho)=2gf^2(\rho)+(1+2g)\rho f(\rho)f'(\rho) \eqnd\eRGmatexact $$
with
$$f(0)=1 \, .$$
Multiplying the equation by $f^{1/2g-1}$, one can integrate and obtains the algebraic equation
$$f^{1/2g}=  \rho f^{1+1/2g}+1\,.$$
{\it Example}: for $g=-1/2$, the equation has the simple solution $$f=1/(1+\rho )\ \Rightarrow\ V(\mu)=\ln(1+\mu^2/2).$$ \par
More generally, for $\rho\to\infty $ and $g<0$, one finds $f\sim 1/\rho\ \Rightarrow\ V(\mu)\sim 2\ln \mu$. 
A first problem with this solution, is that it seems to depend on a {\it continuous parameter}\/ $g$ while we expect a discrete spectrum of multicritical points.\par
Moreover, the behaviour for large matrices is such that the matrix integral no longer converges, potential and measure being of same order. 
\par
We also set $f=1/R$. We then find the fixed point equation
$$RR'+2g R-(1+2g)\rho R'=0\,.\eqnd\ematfixdequation $$
Remarkably enough, we find an equation that is identical to the equation \eqns{\eIRGUflowgen} that appears in vector models and seems to be a {\it universal feature} of this kind of approximations when combined with a local approximation in higher dimensions. We do not need to discuss it again here. 
 \par
In the case of matrix models, for a still improved determination of    critical and multicritical points, one has presumably to introduce more general potentials, that is, {\it with products  of traces},  and  special parametrizations  or   non-linear matrix transformations  as the study of vector models as indicated.
 
%%%%%%%%%%%%%%%%%%%%%%%
\section{Path integral}

We consider now the Euclidean action
$${\cal S}_N({\Bfg\phi})=N\tr \int\d t\, \left[\ud  \dot {\Bfg\phi}^2 (t)  +V\bigl({\Bfg\phi}(t)\bigr)\right],$$
where $V({\Bfg\phi})$ has the   form \eqns{\egenmatpot}.\par

 The problem of quantum mechanics with large size matrices has also been discussed within other RG approaches, in \refs{\rJAPHD,\ \rSDTD}. Field theory has also been considered \refs{\rSNishi}.\par
The discussion now combines the arguments given for the vector path integral with the matrix integral.\par  
 The action ${\cal S}_{N+1}$ expressed in terms the parametrization \eqns{\ephiNun}, becomes
$$ {\cal S}_{N+1}({\Bfg\phi},v)=(1+1/N){\cal S}_N+(N+1)   \int \d t \,\left[\dot v^*(t)\cdot \dot v (t)+ v^* V'({\Bfg\phi}){\Bfg\phi}^{-1}v  \right]+\cdots\ .  $$
After integration over the vector $v$ in the Gaussian approximation, we find
$$ {\cal S}_{N+1}({\Bfg\phi})=(1+1/N){\cal S}_N+\tr\ln\left[-\left({\d \over \d t}\right)^2{\bf 1}+ V'({\Bfg\phi}){\Bfg\phi}^{-1}\right].  $$
We then rescale ${\Bfg\phi}$ and $t$:
$$\eqalign{&{\Bfg\phi}\mapsto {\Bfg\phi}\left[1-(1+\eta)/2N\right], \cr
&t \mapsto t(1-\eta/N),\cr}$$
in such a way that the coefficient of $\dot{\Bfg\phi}^2$ becomes again $\ud$.\par
In the local approximation \eqns{\edetlocal}, the variation of the action takes the form
$$N\delta{\cal S}=\tr\int\d t\,\left[V ({\Bfg\phi} )(1-\eta)-(1+\eta){\Bfg\phi}V'({\Bfg\phi}) +\sqrt{2V'({\Bfg\phi})} \right]. $$
%%%%%%%%%%%%%%%%%%%%%%
This leads to a scalar flow equation involving the function $V(\mu)$:
$$N{\partial V(\mu) \over\partial N}=V(\mu)(1-\eta)-(1+\eta)\mu V'(\mu)+\sqrt{2V'(\mu)}.$$
Differentiating with respect to $\mu$, we obtain
$$ N{\partial V'(\mu) \over\partial N}=-2 \eta V'(\mu) -(1+\eta)\mu V''(\mu)+{V''(\mu)\over\sqrt{2V'(\mu)}}.$$
 and setting
 $V'(x)=1/2R^2 $, one recovers the universal equation \eqns{\eIRGUflowgen}.
 %%%%%%%%%%%%%%%%%%%%%%%%%%%%%%%%%%%%%%%%%%%%%%
\subsection Conclusion

From the analysis of the vector model by RG methods, it seems that at least  two problems have to be handled: {\it non-linear transformations of the integration variables} to generate fixed points with suitable properties  and, for path or field integrals, {\it local expansion} of the contribution to the action of the determinant generated by the partial Gaussian integration. These problems seem to arise also in matrix models. But in matrix models, potentials with product of traces will have to be introduced and this will generate  additional difficulties. \par
Finally, it is quite remarkable that in this RG inspired method, in the large $N$ limit and within the local approximation scheme, {\it only one (one-parameter) fixed point equation arises} both in matrix and different vector models and this point, together with the meaning of the equation, remains to be better understood. 
\listrefs
\bye